\documentclass[
]{elsart}

\usepackage{amsmath,amssymb,graphicx}

\begin{document}

\begin{frontmatter}



\title{Persistent dynamic correlations in self-organized critical
systems away from their critical point}


\author[uaf]{Ryan Woodard\thanksref{now}\corauthref{cor}},
\thanks[now]{Present address: British Antarctic Survey, 
Madingley Road, Cambridge, CB3 0ET, UK}
\corauth[cor]{Corresponding author}
\ead{rywo@bas.ac.uk}
\author[uaf]{David E. Newman},
\author[carlos]{Ra\'{u}l S\'{a}nchez\thanksref{nowornl}},
\thanks[nowornl]{Present address:  Oak Ridge National Laboratory,
Oak Ridge, Tennessee 37831-8070, USA}
\author[ornl]{Benjamin A. Carreras}
\address[uaf]{University of Alaska Fairbanks \\
  Fairbanks, Alaska 99775-5920, USA}
\address[carlos]{Universidad Carlos III de Madrid \\
  28911 Legan\'{e}s, Madrid, Spain}
\address[ornl]{Oak Ridge National Laboratory \\
  Oak Ridge, Tennessee 37831-8070, USA}

\begin{abstract}
  We show that correlated dynamics and long time memory persist in
  self-organized criticality (SOC) systems even when forced away from
  the defined critical point that exists at vanishing drive strength.
  These temporal correlations are found for all levels of external
  forcing as long as the system is not overdriven.  They arise from
  the same physical mechanism that produces the temporal correlations
  found at the vanishing drive limit, namely the memory of past events
  stored in the system profile.  The existence of these correlations
  contradicts the notion that a SOC time series is simply a random
  superposition of events with sizes distributed as a power law, as
  has been suggested by previous studies.
\end{abstract}

\end{frontmatter}


%
%
\newcommand{\newcommandm}[2]{\newcommand{#1}{\ensuremath{#2}}}
\newcommand{\renewcommandm}[2]{\renewcommand{#1}{\ensuremath{#2}}}

\newcommand{\A}{{\em A}}
\newcommand{\B}{{\em B}}
\newcommandm{\LL}{\mathrm{L}}
\newcommandm{\WW}{\mathrm{W}}
\newcommandm{\LLa}{\mathrm{L}_1}
\newcommandm{\LLb}{\mathrm{L}_2}
\newcommandm{\LLL}{\mathrm{L}^2}
\newcommandm{\LLi}{\mathrm{L}^{-1}}
\newcommandm{\alp}{\alpha}
\newcommandm{\acfd}{AC_\mathrm{d}}
\newcommandm{\acfnd}{AC_{\mathrm{nd}}}
\newcommandm{\act}{\rho_a}
\newcommandm{\bb}{\beta}
\newcommandm{\bbg}{\beta_G}
\newcommandm{\bbb}{\beta_B}
\newcommandm{\bbh}{\beta = 2H - 1}
\newcommandm{\bul}{\bullet\;}
\newcommandm{\bulnoin}{\noindent \bullet\;}
\newcommandm{\bulnointwo}{\noindent \hspace{.2in} \bullet\;}
\newcommandm{\bultwo}{\bullet\;}
\newcommandm{\ch}{\chi}
\newcommandm{\cort}{\tau_{1/2}}
\newcommandm{\Dbar}{\overline{D}}
\newcommandm{\dd}{D}
\newcommandm{\F}{f}
\newcommandm{\ff}{1/f}
\newcommandm{\ffi}{f^{-1}}
\newcommandm{\fo}{f^{\,0}}
\newcommandm{\fff}{f^{-2}}
\newcommandm{\ffff}{f^{-4}}
\newcommandm{\fa}{f^{-\alpha}}
\newcommandm{\fb}{f^{-\beta}}
\newcommandm{\fch}{f^{-\chi}}
\newcommand{\fts}{FTS}
\newcommandm{\hx}{H \approx 0.8}
\newcommandm{\hh}{H}
\newcommandm{\ha}{H_{\mathrm{A}}}
\newcommandm{\hc}{H_{\mathrm{C}}}
\newcommandm{\hd}{H_{\mathrm{D}}}
\renewcommandm{\hd}{H}
\newcommandm{\he}{H_{\mathrm{E}}}
\newcommandm{\hf}{H_{\mathrm{F}}}
\newcommand{\hk}{{\em Hwa \& Kardar}}
\newcommandm{\hnn}{h_{\mathrm{n}+1}}
\newcommandm{\hnmn}{h_{\mathrm{n}-1}}
\newcommandm{\hn}{h_{\mathrm{n}}}
\newcommandm{\is}{I_{\mathrm{s}}}
\newcommandm{\je}{J_{\mathrm{E}}}
\newcommandm{\jin}{\overline{J_{\mathrm{IN}}}}
\newcommandm{\jinnaked}{J_{\mathrm{IN}}}
\newcommandm{\jout}{\overline{J_{\mathrm{OUT}}}}
\newcommandm{\joutinst}{J_{\mathrm{OUT}}}
\newcommand{\ltc}{long-time correlations}
\newcommand{\mdia}{$T_{\mathrm{max}}$} 
\newcommandm{\nn}{n}
\newcommandm{\nm}{n - 1}
\newcommandm{\np}{n + 1}
\newcommandm{\nfc}{N_{\mathrm{f}}-critical}
\newcommandm{\nf}{N_{\mathrm{f}}}
\newcommandm{\nft}{N_{\mathrm{f}}/2}
\newcommandm{\nmin}{N_{\mathrm{min}}}
\newcommand{\noneq}{nonequilibrium}
\newcommand{\Noneq}{Nonequilibrium}
\newcommandm{\pdf}{P(i)}
\newcommand{\pdfw}{pdf} 
\newcommandm{\pdfz}{P_Z(i)}
\newcommandm{\mypol}{P_0L}  
\newcommandm{\mypoll}{P_0L^2}  
\newcommandm{\po}{P_0}
\newcommandm{\qbar}{\overline{q}}
\newcommandm{\qt}{q(t)}
\newcommandm{\qto}{q(t_0)}
\newcommand{\raulf}{Ra\'{u}l}
\newcommand{\raul}{Ra\'{u}l S\'{a}nchez}
\newcommand{\raulu}{Ra\'{u}l S\'{a}nchez---Universidad Carlos III de Madrid}
\newcommand{\carlos}{Universidad Carlos III de Madrid}
\newcommandm{\rs}{R/S}
\newcommand{\rsa}{\rs\ analysis}
\newcommandm{\sbar}{\overline{s}}
\newcommand{\sanchez}{S\'{a}nchez}
\newcommandm{\sx}{S(f)}
\newcommandm{\snu}{S(\nu)}
\newcommand{\std}{STD}
\newcommand{\soc}{self-organized criticality}
\newcommandm{\so}{S_0}
\newcommandm{\Ta}{T_{\mathrm{A}}}
\newcommandm{\Tb}{T_{\mathrm{B}}}
\newcommandm{\Tc}{T_{\mathrm{C}}}
\newcommandm{\Td}{T_{\mathrm{D}}}
\newcommandm{\Te}{T_{\mathrm{E}}}
\newcommandm{\Tf}{T_{\mathrm{F}}}
\newcommandm{\Tde}{T_{\mathrm{DE}}} 
\newcommandm{\Tq}{T_{\mathrm{Q}}}
\newcommandm{\Tqbar}{\overline{T_{\mathrm{Q}}}}
\newcommandm{\Ts}{T_{\mathrm{S}}}
\newcommandm{\Tai}{T_{\mathrm{A}}^{-1}}
\newcommandm{\Tbi}{T_{\mathrm{B}}^{-1}}
\newcommandm{\Tci}{T_{\mathrm{C}}^{-1}}
\newcommandm{\Tdi}{T_{\mathrm{D}}^{-1}}
\newcommandm{\Tei}{T_{\mathrm{E}}^{-1}}
\newcommandm{\Tfi}{T_{\mathrm{F}}^{-1}}
\newcommandm{\Tqi}{T_{\mathrm{Q}}^{-1}}
\newcommandm{\Tsi}{T_{\mathrm{S}}^{-1}}
\newcommandm{\Tbar}{\overline{T}}
\newcommandm{\tmax}{T_{\mathrm{max}}}
\newcommandm{\tno}{t_0}
\newcommandm{\ttr}{T_{\mathrm{T}}} 
\newcommandm{\tu}{\tau}
\newcommandm{\uo}{U_0}
\newcommandm{\uoc}{\uo-critical}
\newcommandm{\vg}{V_{\mathrm{g}}}
\newcommandm{\xb}{X_{\mathrm{B}}}
\newcommandm{\xbn}{X_{\mathrm{B}}(n)}
\newcommandm{\xbnm}{X_{\mathrm{B}}(n-1)}
\newcommandm{\xbnp}{X_{\mathrm{B}}(n+1)}
\newcommandm{\xg}{X_{\mathrm{G}}}
\newcommandm{\xgn}{X_{\mathrm{G}}(n)}
\newcommandm{\z}{Z}
\newcommandm{\za}{Z_{\mathrm{A}}}
\newcommandm{\zb}{Z_{\mathrm{B}}}
\newcommandm{\zc}{Z_{\mathrm{crit}}}
\newcommandm{\zcmnf}{Z_{\mathrm{crit}} - \nf}
\newcommandm{\zg}{Z_{\mathrm{g}}}
\newcommandm{\zn}{Z_{\mathrm{n}}}
\newcommandm{\znpo}{Z_{\mathrm{n}+1}}
\newcommandm{\znmo}{Z_{\mathrm{n}-1}}

\newcommand{\mf}[1]{\mathrm{#1}}

\newcommandm{\Apow}{A}
\newcommandm{\Bpow}{B}
\newcommandm{\Cpow}{C}
\newcommandm{\Dpow}{D}
\newcommandm{\Epow}{E}
\newcommandm{\Fpow}{F}
\newcommandm{\Ars}{A}
\newcommandm{\Crs}{C}
\newcommandm{\Drs}{D}
\newcommandm{\Ers}{E}
\newcommandm{\Frs}{F}

\newcommand{\ppow}[2]{$#1 \times 10^{#2}$}

\newcommandm{\trig}{\Delta T_t}
\renewcommandm{\trig}{T_t}
\newcommandm{\quiet}{T_q}
\newcommandm{\dur}{T_d}

\newcommandm{\low}{LD}
\newcommandm{\med}{MD}
\newcommandm{\hi}{HD}
\newcommandm{\zer}{ZD}

\newcommandm{\nq}{\mathrm{nq}}
\newcommandm{\nqt}{N_{\mathrm{qt}}}
\newcommandm{\nnq}{N_{\nq}}
\newcommandm{\ft}{f(\tau)}
\newcommandm{\ftft}{\sum_{\tau = 0}^{N - 1} \left|f(\tau)\right|^2}
\newcommandm{\Fnu}{F(\nu)}
\newcommandm{\Ff}{F(f)}
\newcommandm{\powernaked}{\left|F(\nu)\right|^2}
\newcommandm{\power}{\left|F(\nu)\right|^2}
\newcommandm{\powersum}{\sum_{\nu = 0}^{N - 1} \left|F(\nu)\right|^2}
\newcommandm{\powerf}{\sum_{f = 0}^{N - 1} \left|F(f)\right|^2}
\newcommandm{\powerN}{N \power}
\newcommandm{\ftnq}{f_{\nq}(\tau)}
\newcommandm{\ftnqftnq}{\sum_{\tau = 0}^{N_{\nq} - 1}
  \left|f_{\nq}(\tau)\right|^2}
\newcommandm{\Fnunq}{F_{\nq}(\nu)}
\newcommandm{\powernq}{\sum_{\nu = 0}^{N_{\nq} - 1}
  \left|F_{\nq}(\nu)\right|^2}
\newcommandm{\powernqN}{N_{\nq} \powernq}

\newcommandm{\trefill}{T_{\textrm{refill}}}
\newcommandm{\trefillmin}{T_{\textrm{re,min}}}
\newcommandm{\cmaxdef}{C_{max} = \frac{\zc L^2}{2}}
\newcommandm{\cmindef}{C_{min} = \frac{(\zc - \nf) L^2}{2}}
\newcommandm{\cmaxminuscmin}{C_{max} - C_{min} = \frac{\nf L^2}{2}}
\newcommandm{\fluxin}{\Gamma_{in} = \po L T}
\newcommandm{\fluxinout}{\po L \trefill = \frac{\nf L^2}{2}}
\newcommandm{\trefilldef}{\trefillmin = \frac{\nf L^2}{2 \po L}}
%
%

\newcommand{\onepicht}{2.3in}
\newcommand{\onepicwt}{3in}
\newcommand{\twopicht}{2.in}
\newcommand{\twopicwt}{2.6in}

\section{Introduction}
\label{sec:introduction}

Are events in a self-organized criticality (SOC)~\cite{btw87a,btw88a}
system random or correlated?  This is a still ongoing and important
topic of discussion in the SOC community that has been argued from
both sides to propose or invalidate practical tests for SOC
behavior~\cite{boffetta99a,spada01a,sanchez02a,sanchez02b,naturedebate,sanchez03b,yang04a,woodard04f,yang04b}.
One specific point of this dialogue concerns the distribution of the
amount of time between the initiation of successive events (waiting,
laminar or first return times) in a system suspected of exhibiting SOC
behaviour.  At issue is whether or not the probability density
function (PDF) of this measure should always scale exponentially or as
a power law for a SOC system.  Three recent examples of this are drawn
from space plasma physics~\cite{boffetta99a}, laboratory confined
fusion plasma physics~\cite{spada01a} and earthquake
geophysics~\cite{yang04a}. The authors of all of these studies first
assume that events in a SOC system must be uncorrelated. They then
report observations of waiting times distributed as power laws and
draw separate but similar conclusions that the systems' (space plasma,
fusion plasma, earthquakes) dynamics cannot be consistent with SOC.

The root of this controversy is to be found in the misconceptions that
``strong correlations between successive [events is] at variance with
the SOC model''~\cite{boffetta99a} and that an event ``can occur
randomly anywhere at any time and\ldots cannot `know' how large it
will become''~\cite{naturedebate,yang04a}.  Based on these statements,
the idea of exponentially distributed waiting times could be easily
derived~\cite{boffetta99a}. But it has been known for some time that
these notions simply do not hold in a SOC
system~\cite{christensen92a,sanchez02a}. They were probably
encouraged, in the early days of SOC, by analytical
calculations~\cite{jensen89a,kertesz90a} that appeared to reproduce
quite well the power spectrum of the avalanche activity observed in
the Bak-Tang-Wiesenfeld (BTW) sandpile~\cite{btw87a} (and later in the
directed running sandpile~\cite{hwakardar92a}) when modeling it as a
random superposition of events with power law distributed sizes. Since
then, a state of confusion that still persists appears to surround
this issue, at least among those looking for SOC features in physical
systems, with quite a few papers
accepting~\cite{boffetta99a,spada01a,yang04a,yang04b} and others
denying~\cite{sanchez02a,sanchez03b,woodard04f,davidsen02a} the
validity of these notions.

In this paper we show explicitly that features exist in the power
spectra of SOC systems that signify dynamical correlations and that,
therefore, any assumption or test based on the absence of temporal
correlations is meaningless. By `dynamical' we mean that temporal
correlations are caused by the system dynamics, not by a hypothetical
non-randomness in the external drive~\cite{sanchez02a}. To study them,
we have examined the avalanche activity time series of the same one
dimensional running sandpile model analyzed in
Ref.~\cite{hwakardar92a} as our prototypical SOC model. The benefit of
using this model is that it allows us to analyze systems with a finite
drive strength instead of just the vanishing drive limit in which the
absorbing phase transition responsible for the appearance of
criticality and universal behaviour was
identified~\cite{dickman98a,dickman00a,vespignani00a,pruessner06a}.
Indeed, in many practical applications a finite drive is more
appropriate.  This is the case, for instance, in most magnetized
plasmas, either inside the Sun~\cite{lu95a} or confined in an
earth-based tokamak~\cite{newman96a}.

Taking advantage of increased computer power over that available at
the time of Ref.~\cite{hwakardar92a} we explore timescales up to four
decades longer than those used in that study for varying driving rates
and system sizes. Because of this, dynamical correlations not
previously detected in the power spectrum of the avalanche activity
time series (associated with $f^{-\bb}$ power law regions of exponent
$0 < \bb \leq 1$ ) are now apparent. The exponent is found to increase
with the drive strength, reaching $\bb=1$ only at the higher drives.
We show that this is not a signature of the increasing dominance of
avalanche overlapping (as stated previously in
Ref.~\cite{hwakardar92a}), but just a consequence of the fact that the
timescales where the correlations are present are also displaced
towards higher frequencies at stronger drives. Therefore, avalanche
overlapping is not the cause of the \ff\ region seen in the directed
running sandpile.

The changing value of $\beta$ might seem to suggest that the character
of the correlations is changing with the drive as we move away from
the critical point. However, when analyzing the same activity signals
with rescaled range (\rs) analysis~\cite{hurst51a,mandel2002a} to
estimate the Hurst exponent $H$, a very different conclusion is drawn.
Dynamical correlations are associated with $H > 0.5$.  We find a
\emph{constant} Hurst exponent $\hh \simeq 0.8$ for \emph{all values
  of driving rate} over the same timescales described above, where
$\bb > 0$ in the power spectrum.  $H$ remains constant as \bb\ changes
from $\bb\simeq 0.4$ at the lowest driving rate studied here to
$\bb\simeq 1$ at the highest driving rates. Interestingly, in the
vanishing drive limit~\cite{kadanoff89a} we find the same value of
\hh. In this natural limit of the running sandpile model, avalanches
do not overlap, by definition, because the external drive is turned
off as soon as an event is initiated. The origin of the correlations
must thus be found, then, in a mechanism other than avalanche overlap.
Specifically, the origin of temporal correlations in the sandpile
model is the shape of the system profiles, carved by past avalanches,
where the memory of the system history is stored.

In this report, we do not set out to prove that temporal correlations
exist in a SOC system.  In the words of a helpful referee, ``there is
no question whether correlations exist in self-organized or any other
critical systems.  They do, basically by definition.''  Such
correlations can be proved through the existence of diverging
correlation lengths and times \cite{hinrichsen00a}.  Given that
temporal correlations exist in SOC systems, our goals are to lay to
rest the controversy mentioned in the first paragraph and to address
the questions: How does one identify temporal correlations in a real
system?  What measures should one use?  What do those measures say?
What mechanism(s) produce(s) the correlations?  We use the power
spectrum and \rsa\ in this study but do not claim that they are the
only possible applicable measures.  Multiple measures should be used
for completely thorough studies of dynamical systems.

The paper is organized as follows: in Sec.~\ref{sec:model} the basics
of the running sandpile and analysis techniques are reviewed.
Numerical results obtained for the avalanche activity time series are
described in Sec.~\ref{sec:results} and discussed in
Sec.~\ref{sec:discussion}.  Finally, some conclusions are drawn in
Sec.~\ref{sec:conclusions}.

\section{Model and measures}
\label{sec:model}

\subsection{Model}

The running sandpile can be described by four parameters: \LL, \po,
\zc\ and \nf.  Consider a single column of \LL\ cells.  Each cell
contains an integer number of ``sand grains''; this number is the
height of the cell.  Sand is added to each cell by a random ``rain''
from above.  That is, at each time step for each cell, there is a
probability $0 < \po < 1$ that a grain of sand will be added to it.
The average input current into the entire system is $\jin = \mypol$
grains per time step.  The local gradient \zn\ is the difference in
height between two neighboring cells. If this local gradient exceeds a
critical gradient \zc\ then an avalanche occurs.  An avalanche
stabilizes the local gradient by transferring \nf\ grains of sand from
the higher cell to the lower in a single event called a
flip~\cite{kadanoff89a} or a toppling (the more standard term).  This
avalanche can make \znpo\ and/or \znmo\ unstable at the next time step
so that the avalanche spreads to other cells.  In this way, spatially
and temporally extended events in a system can occur.  This sandpile
can be extended to two or more dimensions~\cite{btw88a} but we will
only discuss the one dimensional case.  For this study, we want to
explore the effects of changing system size (\LL) and external driving
rate (probability \po) so we have arbitrarily fixed the parameters
$\zc = 8$ and $\nf = 3$.  In Ref.~\cite{newman96a} it was shown that
the system dynamics does not qualitatively change for different values
of \zc\ and \nf, as long as $1<\nf<\zc/2$.

Note that time is well-defined in this running sandpile model as
opposed to the vanishing drive model (the original BTW model, for
instance) where external forcing (sand addition) is suspended as long
as there is activity in the system.  In this running sandpile, sand
can be added at each cell at every time step with a probability \po\
even if an avalanche is active.  When we increase the level of
external forcing by increasing \po, we increase the probabilty of a
grain of sand being added to each cell and therefore increase the
probabilty of an avalanche.  But we do not change the time step---the
clock continues to tick regularly regardless of system activity.
Computationally, this requires simultaneous random adding of grains
(forcing) and checking for instability (relaxation) within the same
programming loop.  A natural time scale for the system, then, can be
thought of as $\po^{-1}$.

The time series that we analyze is the instantaneous total avalanche
activity or, in short, the \emph{activity} \act, which is defined as
the total number of unstable sites (where $Z \geq \zc$) at each time
step.  An unstable cell transports \nf\ grains of sand to the next
cell.  The activity at each time step can be thought of as the
instantaneous (potential) energy dissipation in the system or the
total instantaneous flux through it.  Avalanches in a \act\ time
series appear as structures---sequential nonzero activity separated by
periods of inactivity, or quiet times.  We analyze activity data for
times after steady state has been reached, which is when the long time
average number of topplings approaches a constant value.

Two temporal measures from the activity series are discussed in this
paper: durations and quiet times. A duration is the number of time
steps during which the system is active; we refer to this as an
avalanche.  Note that for the running sandpile that we use in this
study, an avalanche defined in this way could actually comprise two or
more distinct avalanches that happen to be active at the same time in
the system.  At high drive, this tends to be the case but at low drive
individual avalanches can be distinguished through analysis of the
pulse shape in the activity time series.  A quiet time is defined as
the length of an uninterrupted sequence of zero topplings (inactivity)
between two avalanches.  It is the length of time that the system is
in an absorbing state.  In the randomly driven sandpile, probability
density functions (PDFs) of unadulterated avalanche durations scale as
a power law in the self-similar range~\cite{jensen98a} while the PDFs
of quiet times scale exponentially~\cite{sanchez02b,sanchez02a}.

An important part of this study is to quantify the effect of changing
the drive strength and system size on correlations.  First, it is
important to note that for critical dynamics to exist, the system must
not only be underdriven (i.e., the flux in, $\jin=P_0L$, must never
exceed the maximum possible flux out of the bottom cell, $\jout=\nf$),
but it must also satisfy $\jin<\nft$, so that a cell can alternate
between stable and unstable until the avalanche ends or washes past
the cell~\cite{newman96a}. Second, note that the difference between
driving close/far from the condition $\jin < \nft$ is to
increase/decrease the degree of avalanche overlapping.  That is, if
$\jin \ll \nft$ then there will be very few cases of simultaneous
avalanches and hence very little avalanche overlapping.  In practise,
\jin\ can be set as small as desired to effectively eliminate all
overlapping, thus allowing identification and analysis of single
avalanches.  In this limit, avalanche dynamics and statistics are the
same as the vanishing drive (non-running) sandpile of
\cite{kadanoff89a}.

\begin{figure*}
  \centering
  \begin{tabular}{cc}
    \includegraphics[width=\twopicwt]{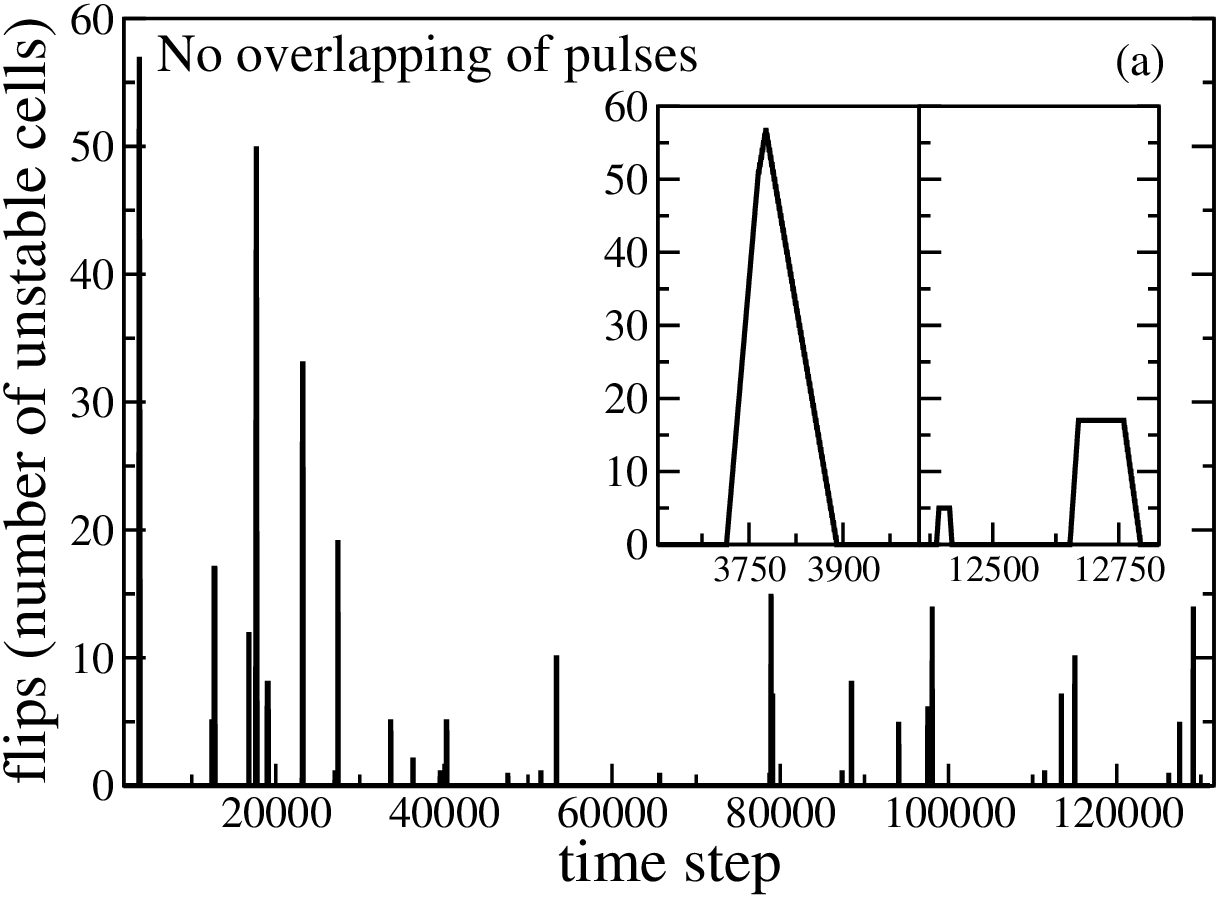} &
    \includegraphics[width=\twopicwt]{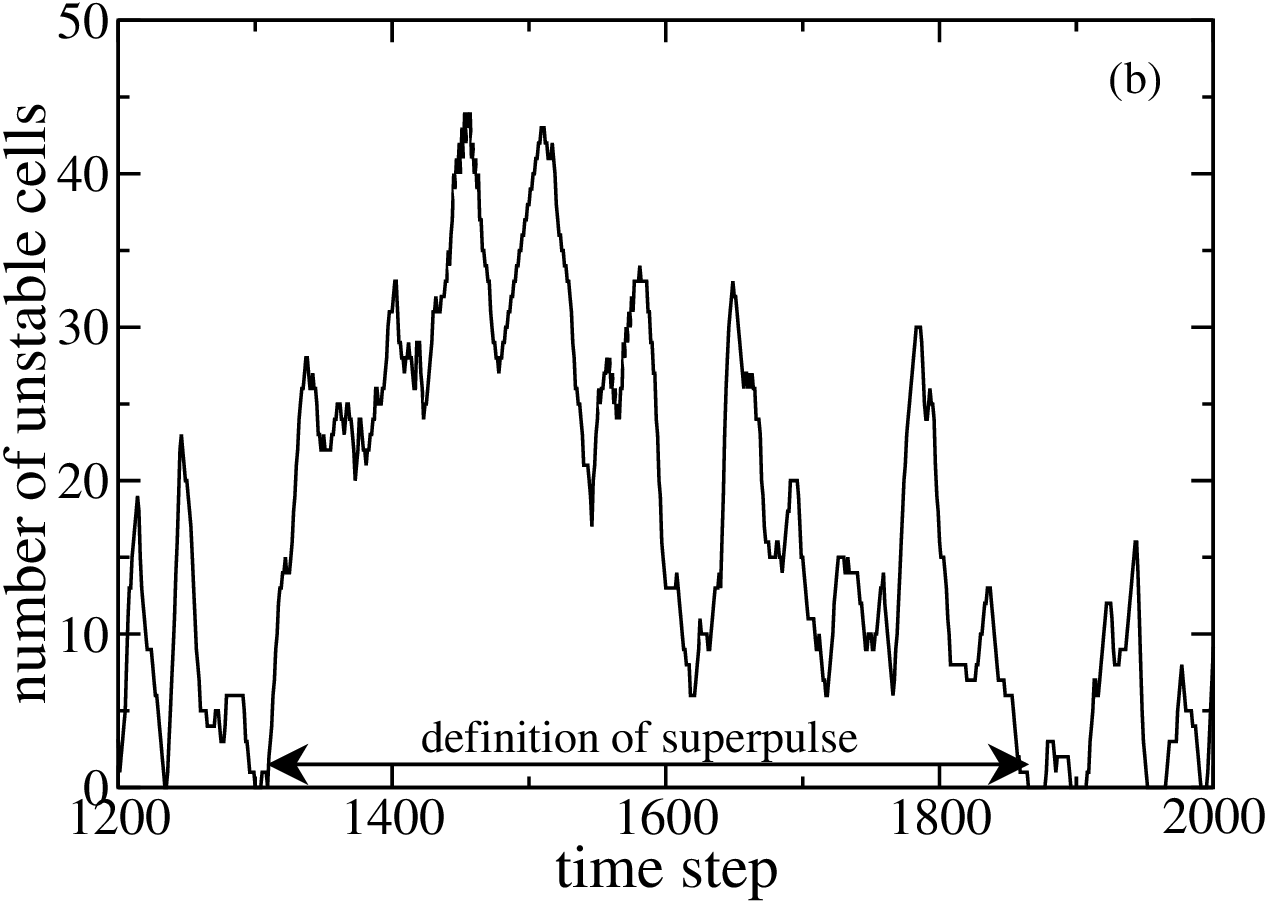}
  \end{tabular}
  \caption[Partial activity time series from running sandpile model at
  steady state for (a) low and (b) high drive.]  {Partial activity time
  series from running sandpile model at steady state for (a) low and
  (b) high drive.  At low drive ($\mypoll = 0.2$), there are
  approximately 30 distinct and separate events in $\sim 10^5$ time
  steps while at high drive ($\mypoll = 100$) there are far more events
  and their identification is made difficult by almost constant
  overlapping.  Note the trapezoid shape of a single, distinct
  avalanche in the magnified inset of the low drive case.  A
  superpulse is defined for high drive as the sequence of activity
  between consecutive quiet times.}
  \label{fig:timeserieslowhigh}
\end{figure*}

We want to quantify the amount of temporal event overlapping and we
want to compare systems of different sizes (\LL).  If the average
avalanche duration $\bar d~(\sim L)$ is greater than the average
trigger time $\trig ~(\sim \nf / \mypol)$~\cite{sanchez02a} then
avalanches will overlap and therefore the condition for overlapping in
time is $\bar d > \trig$ or $\mypoll > \nf$.  (This is in contrast to
the limit for spatial overlap previously given in Eq. (4)
in~\cite{hwakardar92a}.)  For a fixed $N_f$, this condition
encompasses both size and drive strength so we define the {\em
  effective driving rate} as $\je = \mypoll$.  At steady state, of
course, flux into and out of the system are still equal, $\jin =
\mypol$.  \je\ is simply a parameter defined to compare systems of
different sizes that have proportional levels of overlapping.  In what
follows, \emph{low drive} will refer to systems where overlapping of
events is negligible (i.e., $\je \ll \nf$), and \emph{high drive} will
refer to cases when overlapping becomes significant (i.e., $\je \gg
\nf$) (see Fig.~\ref{fig:timeserieslowhigh}).  This will be discussed
further in Section~\ref{sec:size-drive}.

\subsection{Measures}

Though a variety of measures can be used, our tools to detect
correlations are power spectra and rescaled range (\rs) analysis.  The
discrete power spectrum of a data set $X(t)$ is defined as the square
of the discrete Fourier transform, $S(f) = \left|F(f)\right|^2$, where
\begin{equation}
  F(f) = N^{-1} \sum_{t = 0}^{N - 1} X(t) e^{-i 2 \pi (f / N) t}.
\end{equation}
What does the power spectrum say about correlations in a series?  More
appropriate here, what does a power spectrum that scales as a power
law $f^{-\beta}$ have to say about persistence in a series?  For
$\beta > 0$, the low frequencies are more important in the dynamics so
the series tends to look relatively smooth and trends persist.  Such
series are said show persistence. For $\beta < 0$, the high frequency
components are more important and the series is very rough.  Trends do
not persist for long and the series is called antipersistent.  Truly
random noise (white noise) has a spectrum that scales with $\beta =
0$~\cite{bracewell}.

The Hurst exponent, $H \in [0, 1]$~\cite{hurst51a}, can be used to
measure persistence in a time series.  A value of $H > 0.5$ implies
persistence, $H < 0.5$ antipersistence and $H = 0.5$ uncorrelated.  We
measure $H$ by using $R/S$
analysis~\cite{hurst51a,mandel1999a,mandel2002a}.  The rescaled range
is defined as $R'(\tau) \equiv R(\tau)/S(\tau)$, where $S(\tau)$ is
the standard deviation and
\begin{gather*}
  R(\tau) = \underset{1 \leq k \leq \tau}{\max}W(k,\tau) - \underset{1
    \leq k
    \leq \tau}{\min}W(k,\tau) \; \; \; \mathrm{(range),} \\
  W(k,\tau) = \sum_{t=1}^{k} (X_t - \langle X \rangle_\tau) \; \;
  \mathrm{(cumulative \; \; deviation) \; \; and} \\
  \langle X \rangle_\tau = \frac{1}{\tau} \sum_{t=1}^{\tau} X_t \; \;
  \; \mathrm{(mean).}
\end{gather*}
If the rescaled range of the time series exhibits regions that scale
as $R'(\tau) \sim \tau^H$, then $H$ is the Hurst exponent.

\textbf{Need to work on this paragraph.} In some cases, \hh\ as
estimated by \rsa\ can be related to \bb.  For instance, for a time
series of stationary independent identically distributed random
variables (drawn from a Gaussian distribution), \bbh~\cite{mandel82a}.
If, however, the random variables are instead distributed with a fat
tail that scales as $\sim x^{-\mu}$ then the two parameters are more
generally related as $\beta = 2(H + 1 - \mu^{-1})$~\cite{mandel2002a}.
In both examples, a crucial underlying assumption is that the random
variables that comprise the time series are stable and stationary.
This should be kept in mind for Section \ref{sec:partI_region-d:-soc}
below, where we show and discuss the fact that \bb\ and \hh\ are not
simply related for the activity of the running sandpile.

\section{Results}
\label{sec:results}

In this section, we describe the primary qualitative features of the
power spectra and \rsa\ of activity time series obtained for different
driving rates and system sizes.  The main theme is that distinct power
law regions always exist in both measures; we defer the
interpretations of the regions to the following section.

\begin{figure}
  \centering
  \includegraphics[height=5.5in]{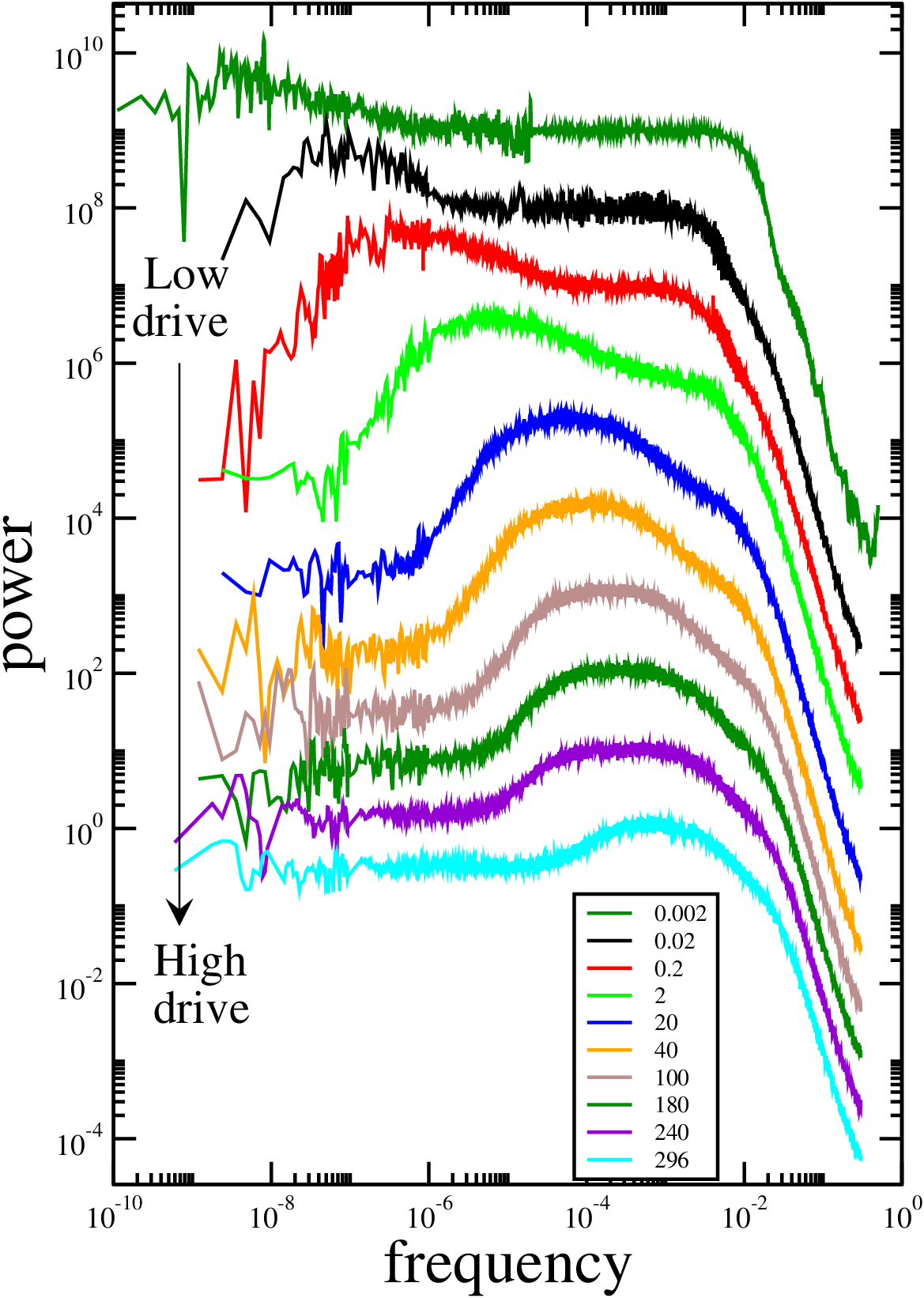}
  \caption[(Color online.) Power spectra of activity time series of
  $\LL = 200$ sandpile for five orders of magnitude of effective
  driving rate in $\mypoll \in \[0.002, 296\]$.]{(Color online.) Power
    spectra of activity time series of $\LL = 200$ sandpile for five
    orders of magnitude of effective driving rate in $\mypoll \in
    [0.002, 296]$.  Spectra have been shifted along the $y$ axis for
    easier viewing.}
  \label{fig:merged_all_stretched}
\end{figure}

The power spectra of the activity look very different depending upon
whether the driving rate is low or high.
Fig.~\ref{fig:merged_all_stretched} shows the spectra for over five
orders of magnitude of effective driving rate, $\je = \mypoll$, which
increases from top to bottom in the figure. [The sandpile size is $\LL
= 200$, but very similar results have been obtained for sizes between
$\LL = 50$ and $\LL = 2000$.] The lowest drive used is $\mypoll =
0.002$ and the highest is $\mypoll = 296$, chosen to stay below the
normal overdrive limit of $\mypol < \nf / 2$, discussed above.
Multiple distinct power law regions are seen in all of the spectra.
The highest frequency regions share a common slope ($\simeq 3.5$). At
low drive a very prominent hump at low frequency moves to higher
frequency as driving rate increases.

\begin{figure*}
  \centering
  \begin{tabular}{cc}
    \includegraphics[width=\twopicwt]{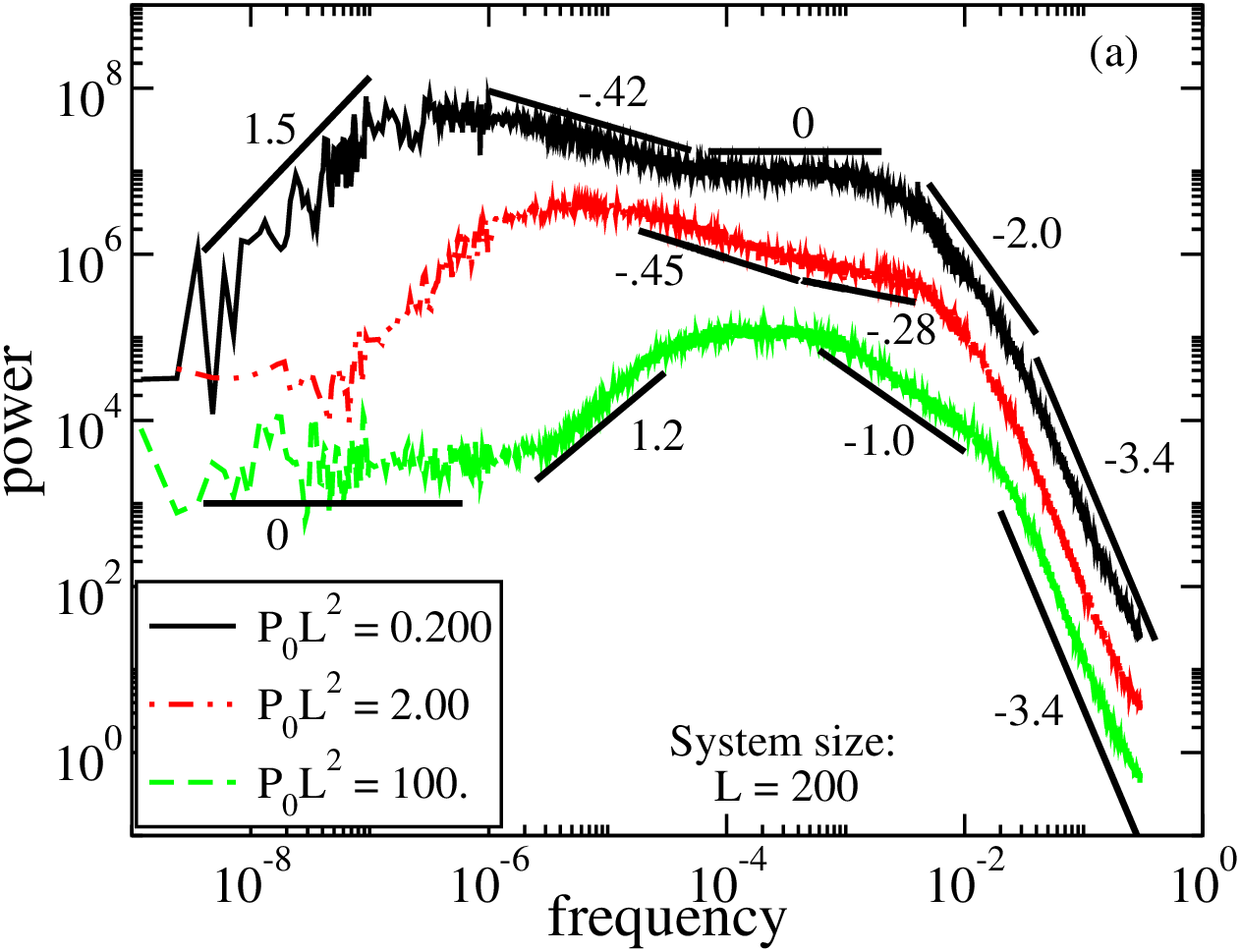} &
    \includegraphics[width=\twopicwt]{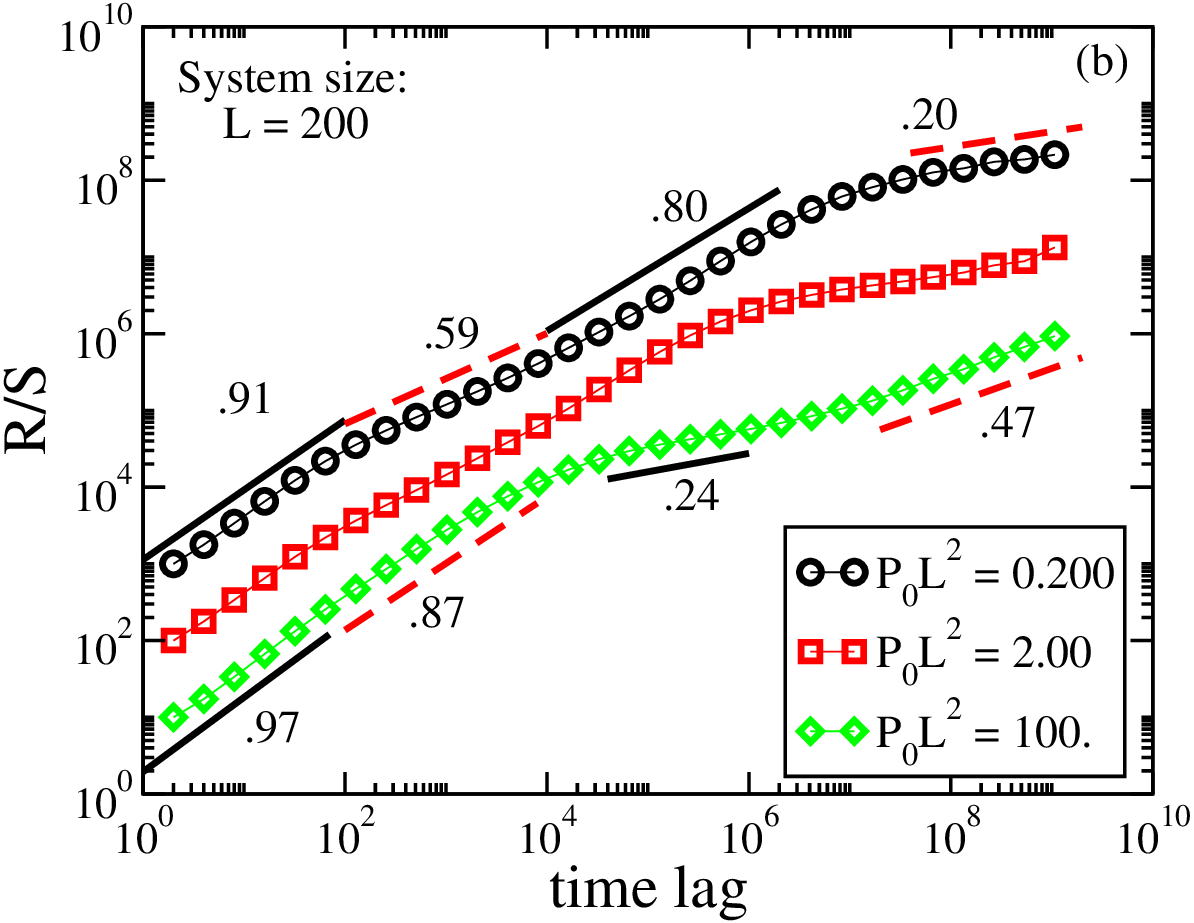} \\
  \end{tabular}
  \caption[(Color online.) (a) Power spectra and (b) \rsa\ of activity
  for fixed system size and low, medium and high driving rates.]
  {(Color online.) (a) Power spectra and (b) \rsa\ of activity for
    fixed system size and low, medium and high driving rates.  The $y$
    values of both measures have been shifted for easier viewing.  In
    the spectra (\rs), six (five) regions of low drive and four
    regions of high drive are shown by solid lines.  Lines are power
    laws; numbers show values of \bb\ in spectra and \hh\ in \rs.
    Lowest frequency \fo\ region of spectra and $\hh = 0.5$ of \rs\
    for the low drive case is not seen because of the finite size of
    the time series.  Its existence is assumed based on the \fo\
    regions seen in the spectra of higher drive cases.}
  \label{fig:powerrs}
\end{figure*}

\begin{figure*}
  \centering
  \begin{tabular}{cc}
    \includegraphics[width=\twopicwt]{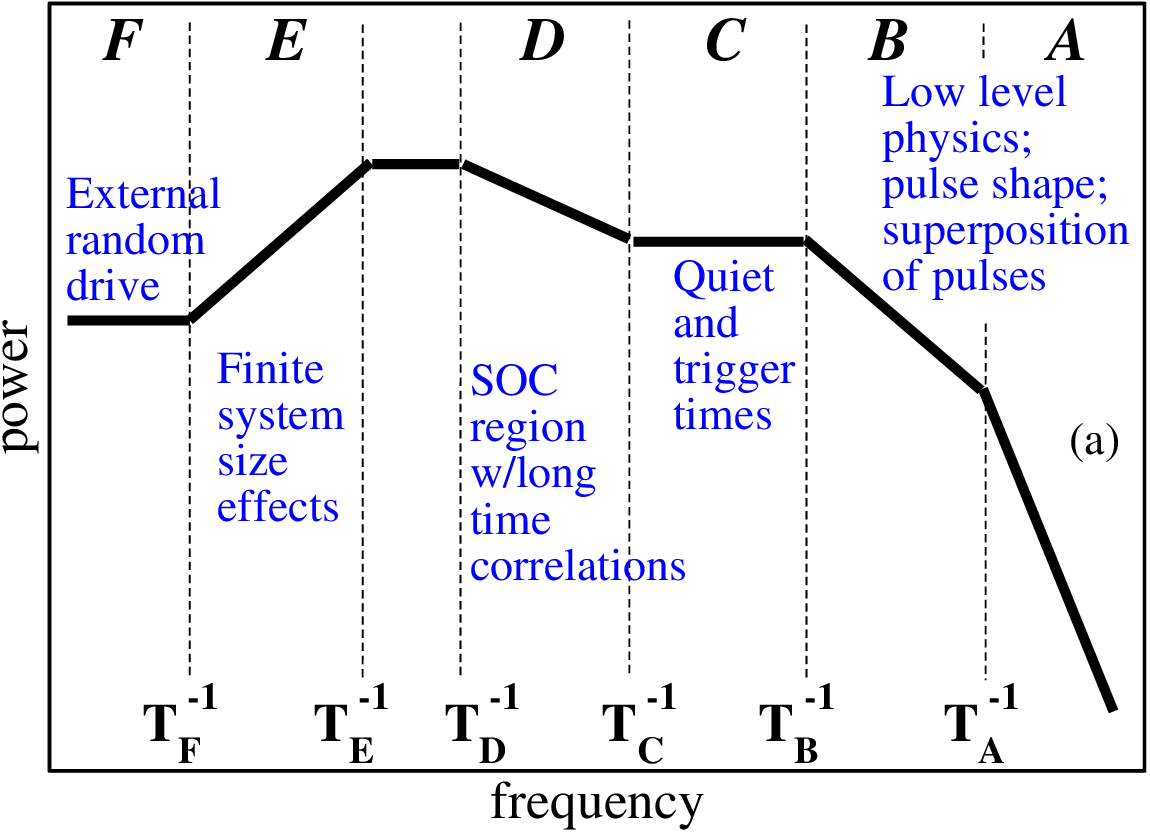} &
    \includegraphics[width=\twopicwt]{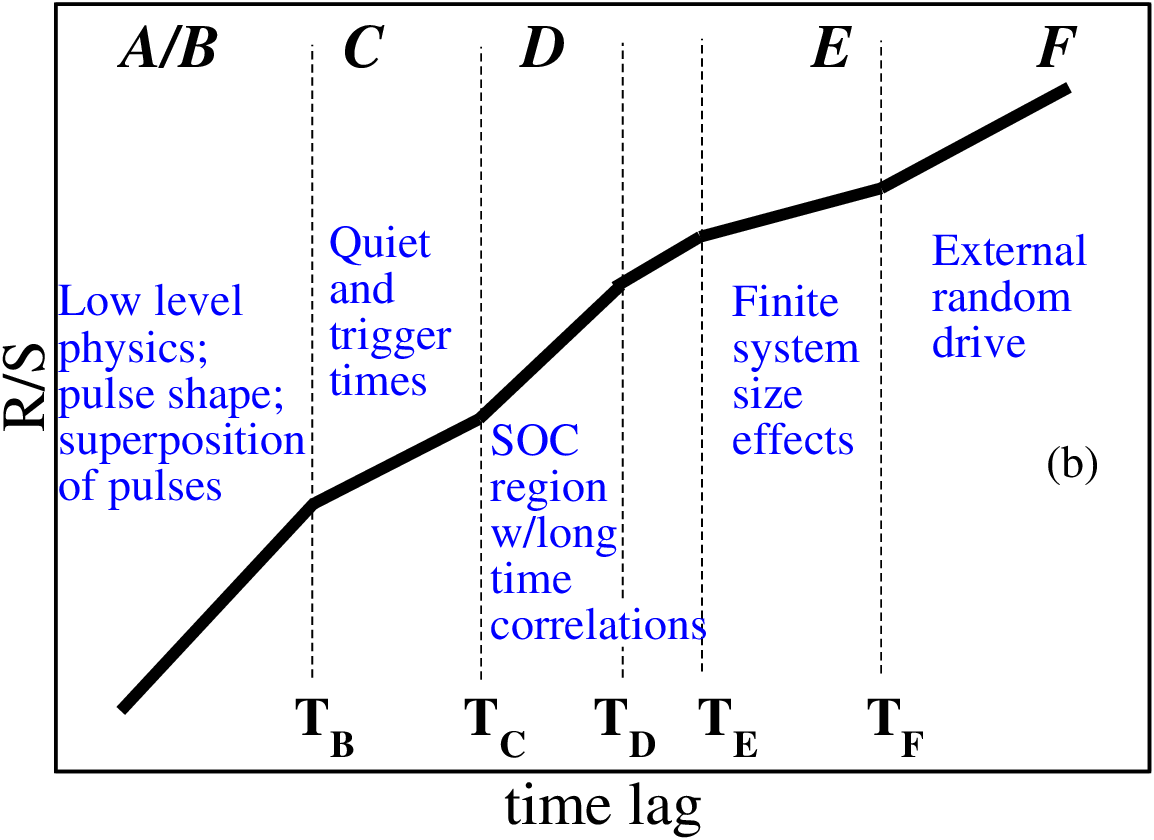} \\
    \includegraphics[width=\twopicwt]{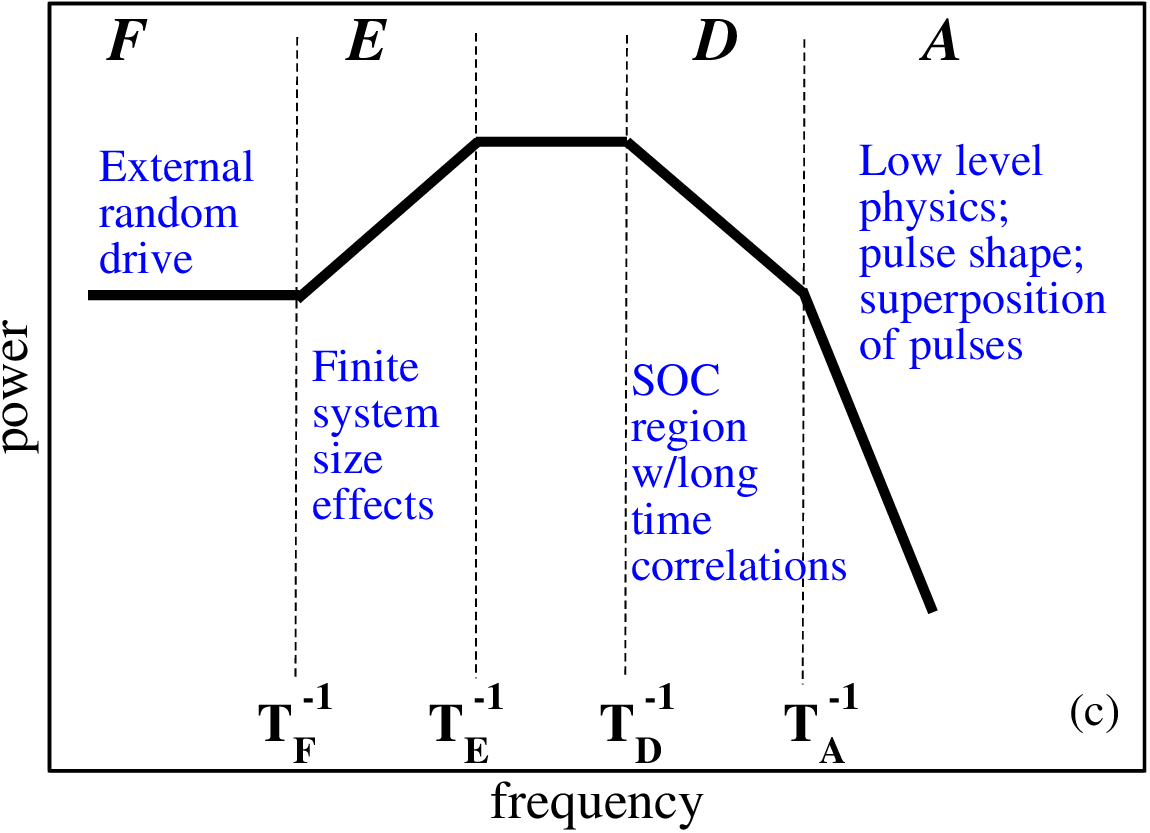} &
    \includegraphics[width=\twopicwt]{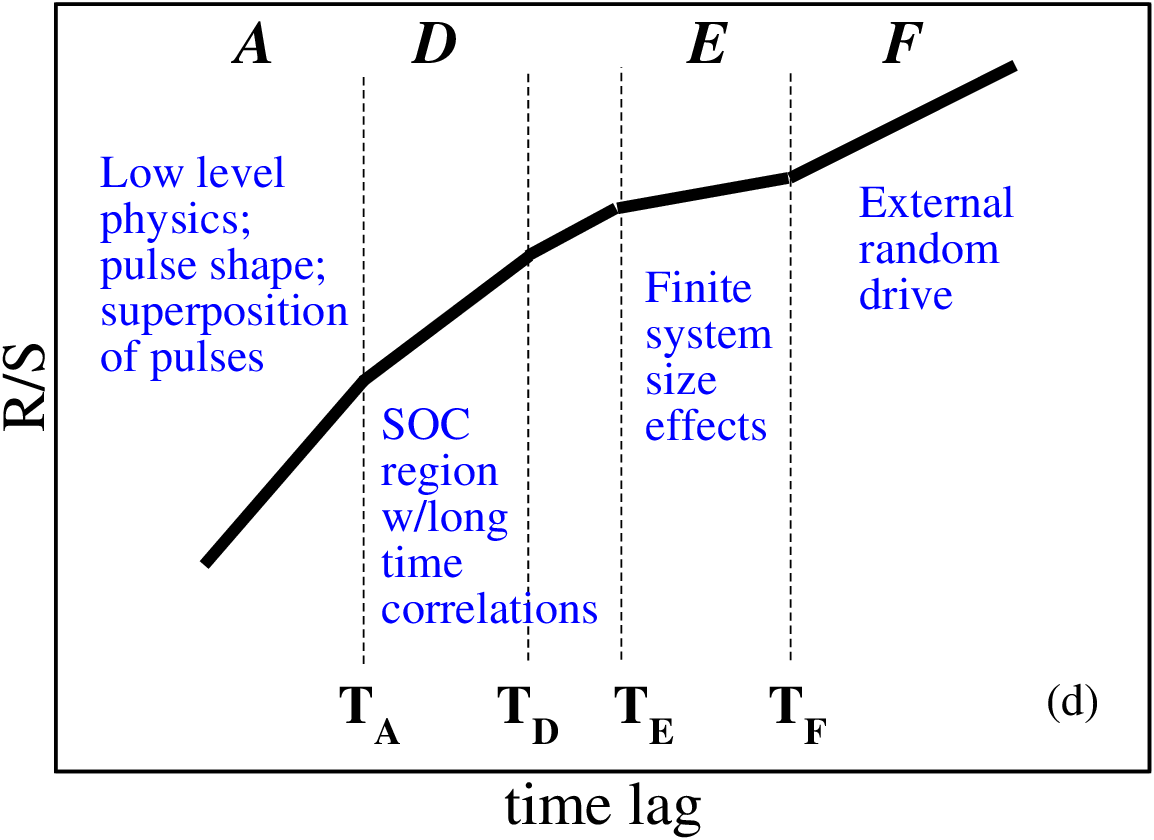} \\
  \end{tabular}
  \caption[Cartoons of distinct regions and their breakpoints and
  causes of power spectra and \rsa\ of sandpile activity.]{Cartoons of
    distinct regions and their breakpoints and causes of power spectra
    and \rsa\ of sandpile activity.  (a) Power spectrum of low drive,
    (b) \rsa\ of low drive, (c) power spectrum of high drive and (d)
    \rsa\ of high drive.  (c) is taken from Fig. 6
    of~\cite{hwakardar92a} and the others are drawn in that spirit.}
  \label{fig:cartoons}
\end{figure*}

For clarity, selected spectra for a low, a medium and a high drive
case are shown in Fig.~\ref{fig:powerrs}(a) for $L=200$.  For the
lowest drive case, six power-law regions can be distinguished, labeled
$A$ to $F$ as frequency decreases. They are separated by breakpoints
denoted by $T^{-1}_k$, $k$ being the appropriate region label.
Similarly, four regions are found in the highest drive case, labeled
$A$, $D$, $E$ and $F$ for reasons to be discussed in the next section.
The regions and the labelling convention are illustrated in the
cartoons in Figs.~\ref{fig:cartoons}(a) and (c).
[Fig.~\ref{fig:cartoons}(c) is based on Fig. 6 of
Ref.~\cite{hwakardar92a}.]

Next, we will describe the \rsa\ results. They are shown, for the same
selected cases, in Fig.~\ref{fig:powerrs}(b). Again, different power
law regions can be distinguished: five distinct regions for the lowest
drive case and four for the highest. They are labeled according to the
same convention as the power spectra (Figs.~\ref{fig:cartoons}(b) and
(d)), except for the shortest timescale region, labeled instead as
``A/B''. Again, relevant breakpoints are identified and labeled.

\begin{figure*}
  \centering
  \begin{tabular}{cc}
    \includegraphics[width=\twopicwt]{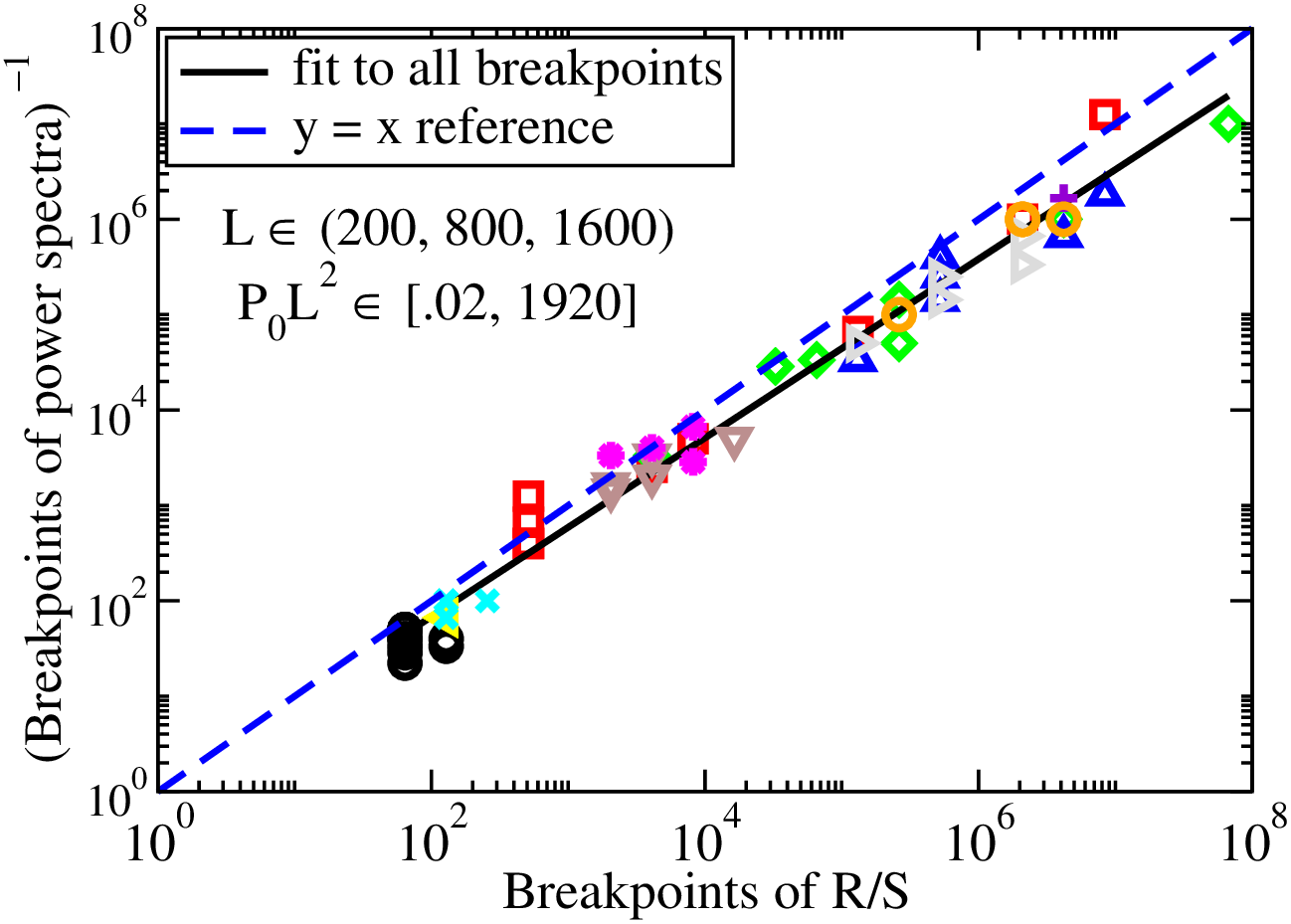} &
    \includegraphics[width=\twopicwt]{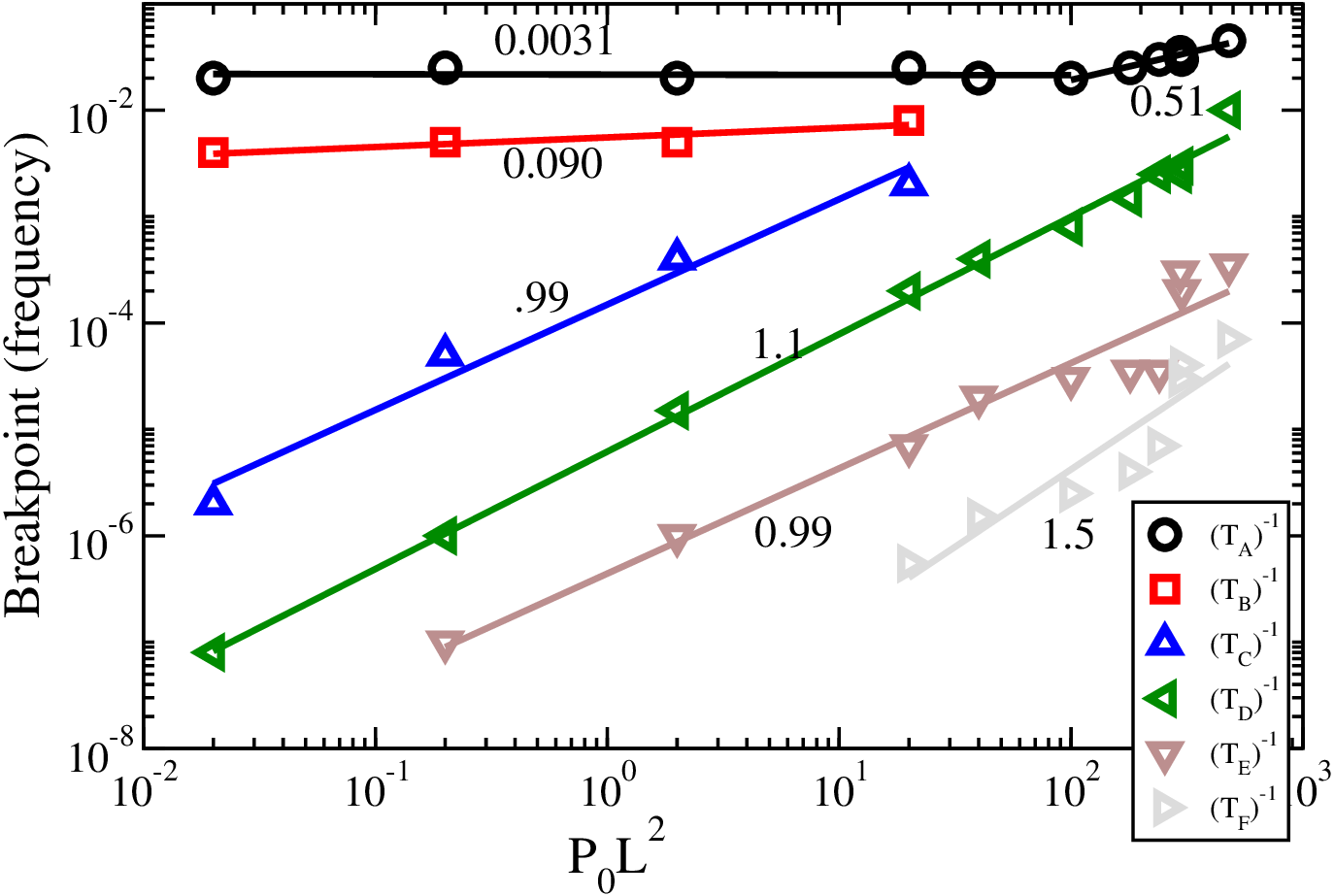}
  \end{tabular}
  \caption[(Color online.) (a) Inverse breakpoints of the power
  spectra versus breakpoints of \rsa\ of activity for different size
  sandpiles and driving rate and (b) breakpoints of power spectra
  versus driving rate for $\LL = 200$ sandpile.]{(Color online.) (a)
    Inverse breakpoints of the power spectra versus breakpoints of
    \rsa\ of activity for different size sandpiles and driving rate
    and (b) breakpoints of power spectra versus driving rate for $\LL
    = 200$ sandpile.  Numbers shown are exponents of power law fits to
    the data.}
  \label{fig:breakpoints}
\end{figure*}

The labelling chosen for the spectrum and \rsa\ breakpoints is not
arbitrary. Breakpoints are compared in Fig.~\ref{fig:breakpoints}(a),
where it is found that those of the two measures, found independently,
agree very closely with each other.  [The \rs\ breakpoints appear at
slightly longer timescales than those of the power spectrum, but this
effect is known from comparisons with Hurst exponents determined via
the structure functions method~\cite{gilmore02a,gilmore03a}.] We
conclude that both measures distinguish the same dynamical regions
through the identification of different power law regions.

For the discussion in the next section it is interesting to know how
these breakpoints scale with driving rate (see
Fig.~\ref{fig:breakpoints}(b)). Two different behaviors can be
distinguished: the breakpoints associated with the $A$ and $B$ (and
$A/B$) regions are independent of the drive, while all of the others
scale (almost) linearly with it, moving to higher frequencies (or
shorter time lags for the \rsa) as the drive increases.

\begin{figure*}
  \centering
  \begin{tabular}{cc}
    \includegraphics[width=\twopicwt]{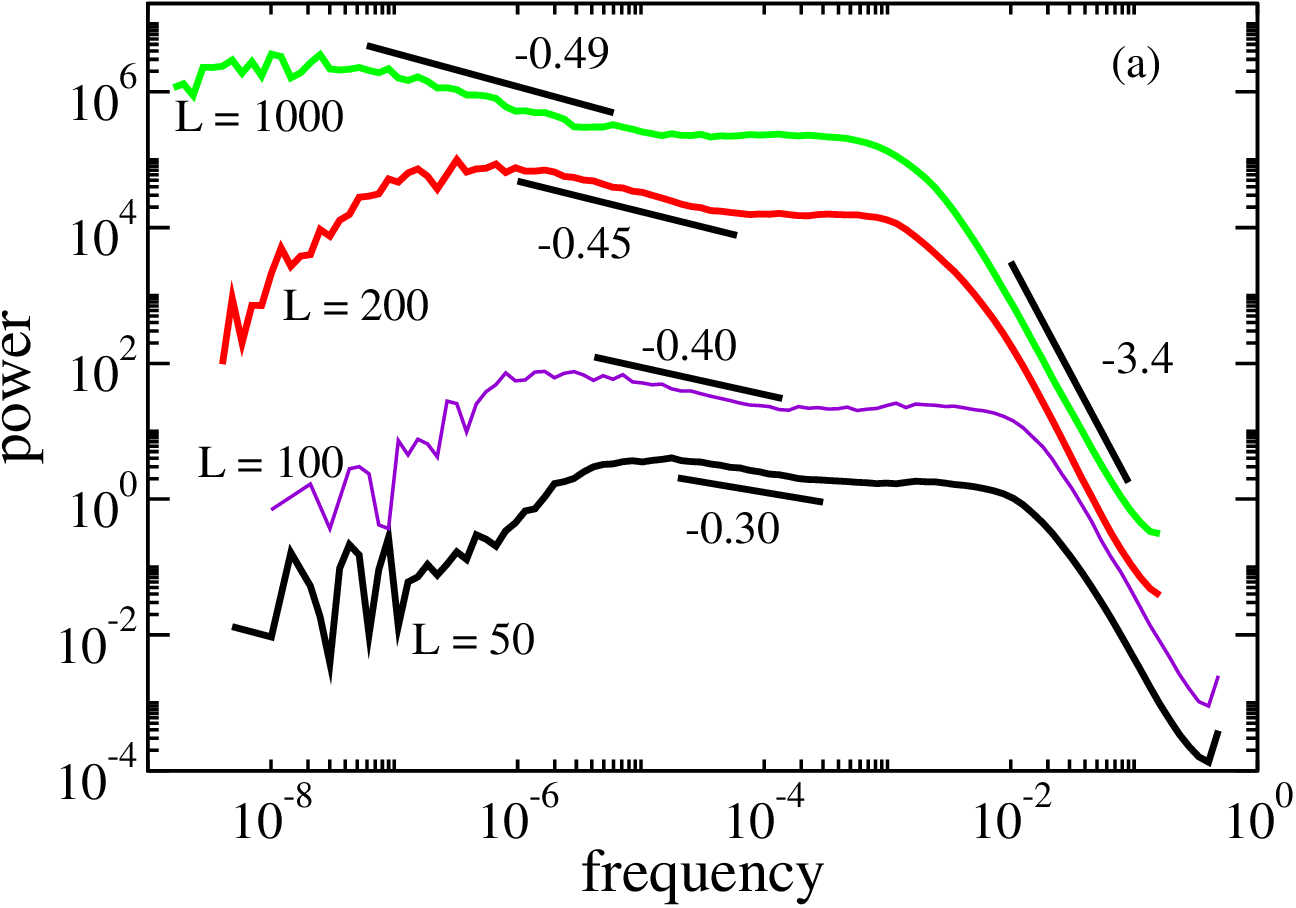} &
    \includegraphics[width=\twopicwt]{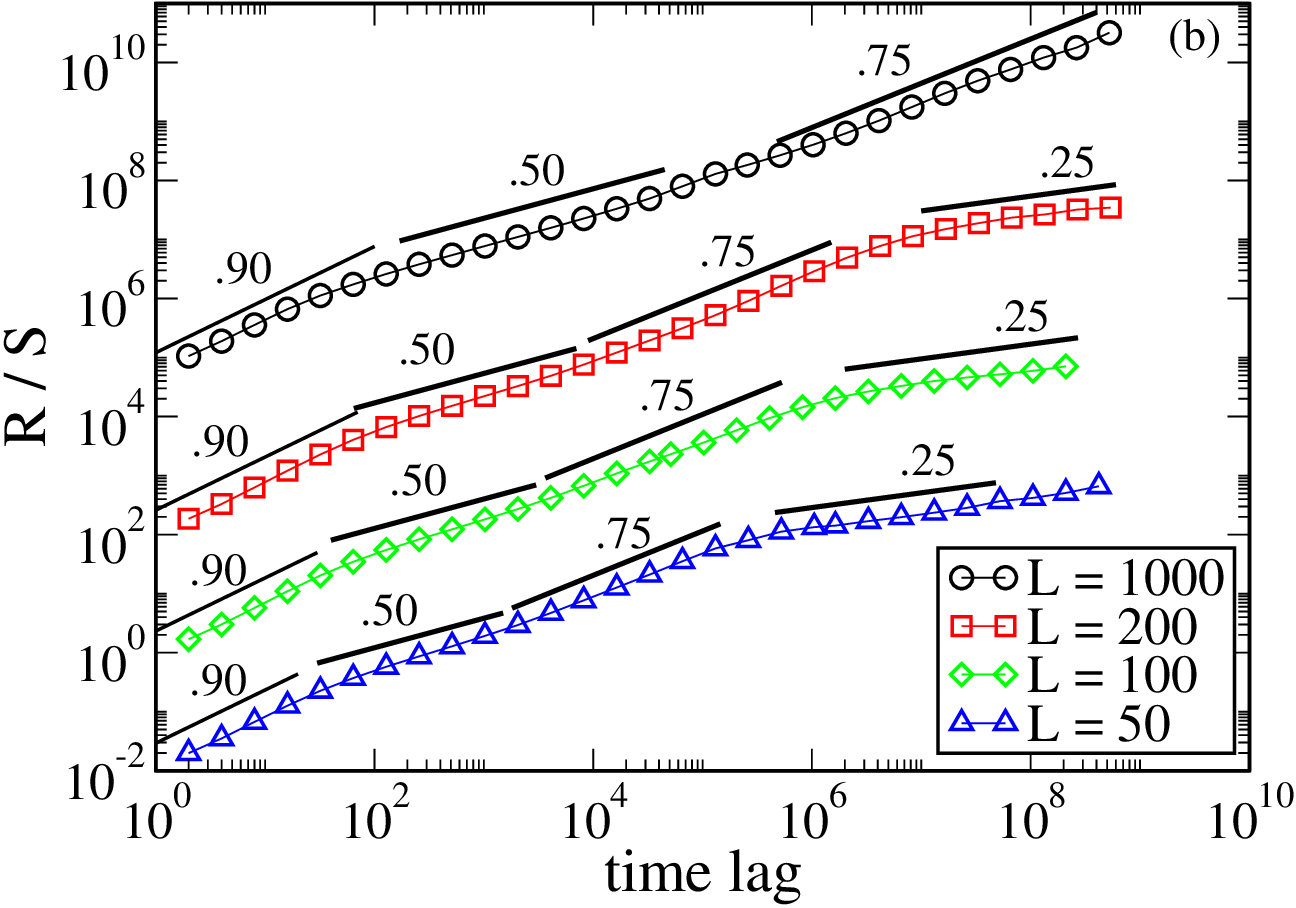}
  \end{tabular}
  \caption[Power spectra and $R/S$ analysis of activity for low drive
  sandpile for $P_0L^2 = 0.2$ and $\mathrm{L} = 20, (ADD 50, 100) 200,
  1000$.]  {Power spectra and $R/S$ analysis of activity for low drive
    sandpile for $P_0L^2 = 0.2$ and $\mathrm{L} = 50, 100, 200, 1000$.
    Plots have been shifted along the $y$ axes for better viewing.
    The solid lines are power laws plotted with the labeled exponent;
    they are not fits to the data.  These plots show how the distinct
    regions move to longer time scales with larger system size but
    with the same effective driving rate, \mypoll.}
  \label{fig:flipsdiffLpowerrs}
\end{figure*}

To conclude the enumeration of results, we will briefly describe the
effect that changing the sandpile size has on the signatures and then
more fully discuss this effect in Section \ref{sec:size-drive}.  If
$L$ is increased while \mypoll\ is held constant, then all regions
move to smaller frequencies (or longer time lags),
Fig.~\ref{fig:flipsdiffLpowerrs}. The values of the exponents
associated with each distinct region remain essentially unchanged,
especially in the \rsa.

\section{Discussion}
\label{sec:discussion}

\renewcommand{\cap}{Power spectra and $R/S$ analysis of low drive
  sandpile: original series, with quiet times removed and with pulses
  shuffled.}
\begin{figure*}
  \centering
  \begin{tabular}{cc}
    \includegraphics[width=\twopicwt]{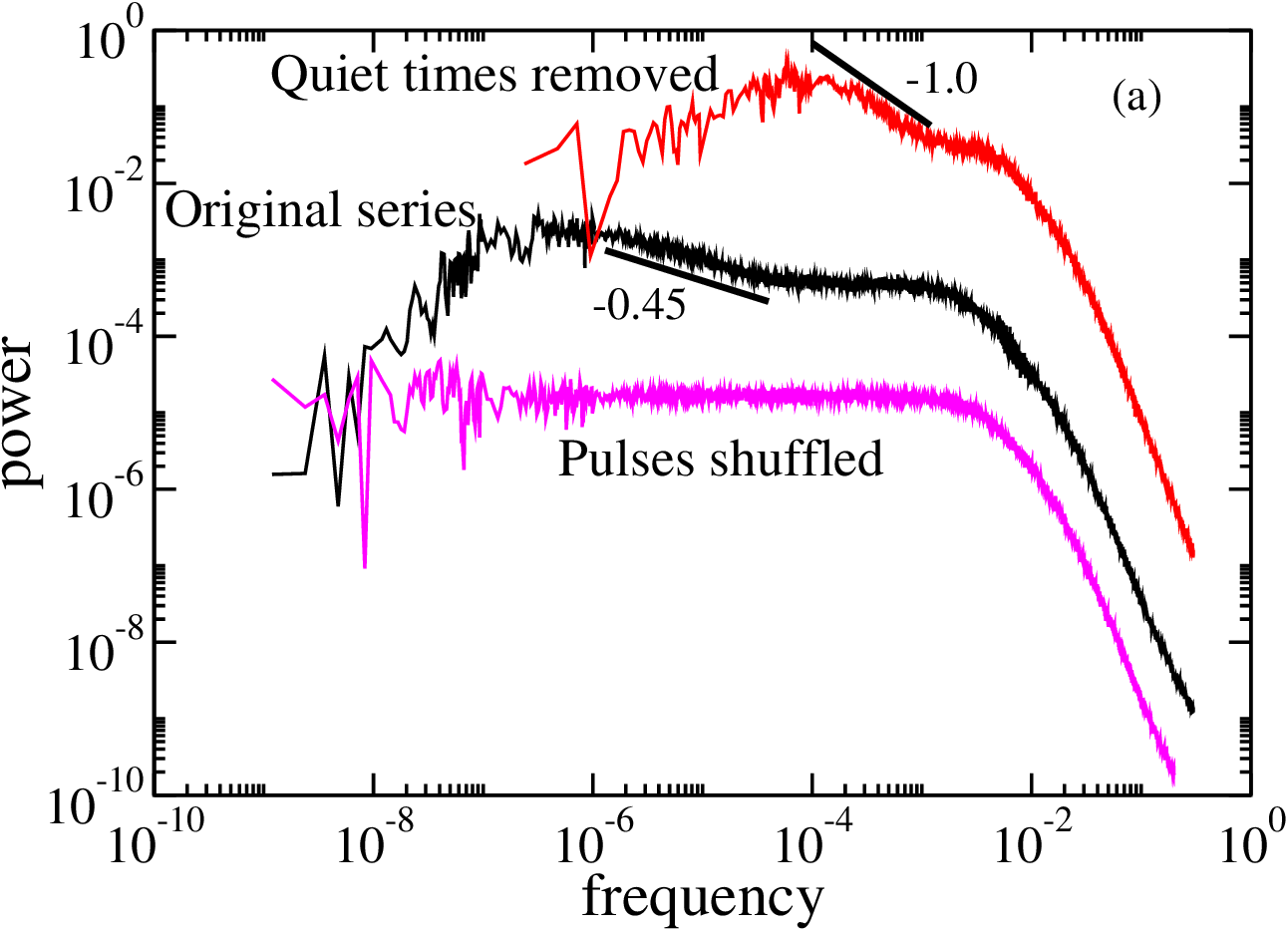} &
    \includegraphics[width=\twopicwt]{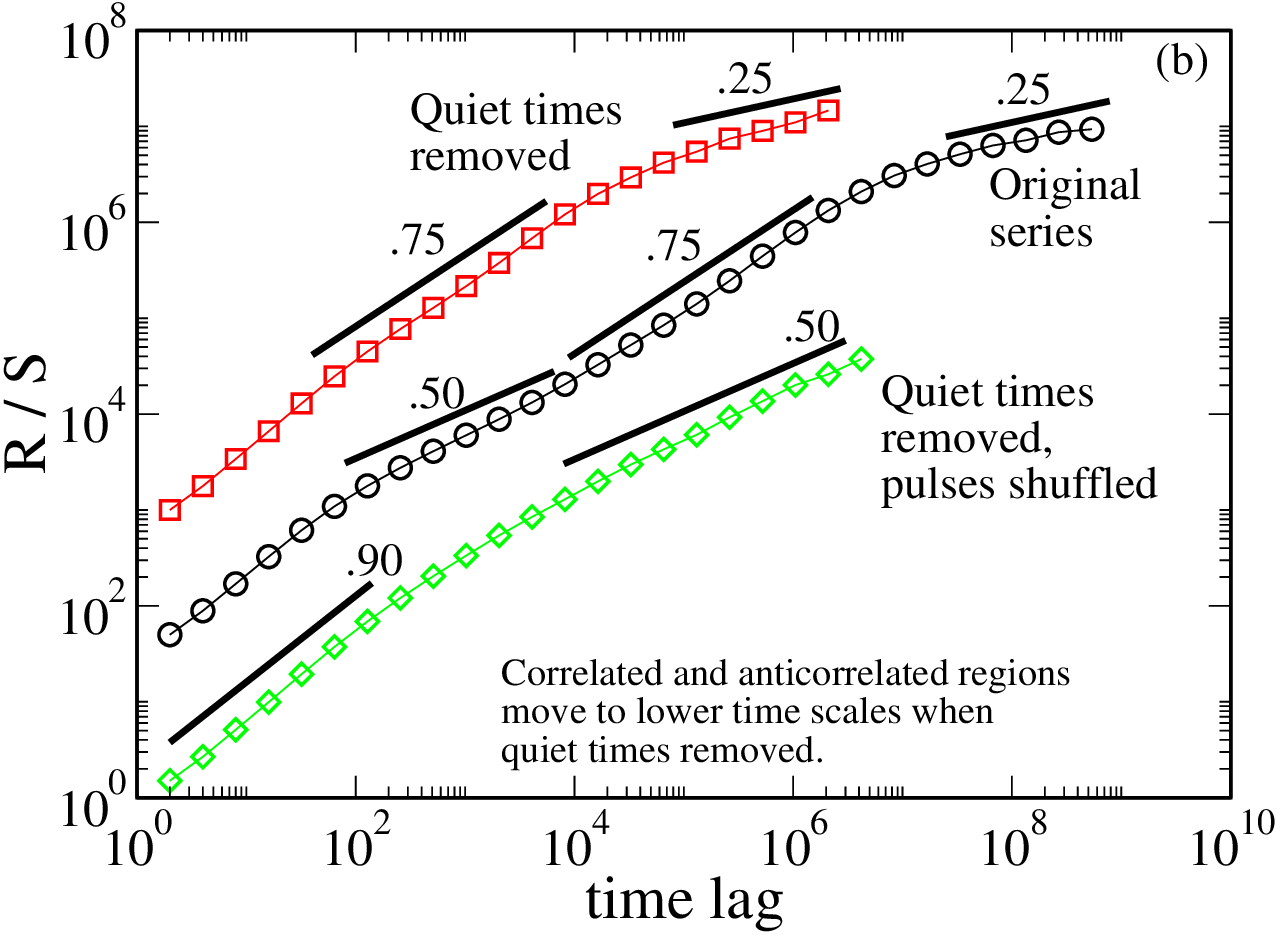}
  \end{tabular}
  \caption[\cap]{\cap\ Plots have been shifted along the $y$ axes for
    better viewing.  The solid lines are power laws plotted with the
    labeled exponent; they are not fits to the data.  The spectrum and
    \rs\ of the activity time series for the zero drive sandpile are
    virtually identical to the spectrum and \rs\ of the low drive
    series with quiet times removed; to preserve clarity, they are not
    shown.}
  \label{fig:powerrsorigshufnq}
\end{figure*}

The first point that can be made from
Figs.~\ref{fig:merged_all_stretched}, \ref{fig:powerrs} and
\ref{fig:flipsdiffLpowerrs} is that long time correlations {\em do}
exist in the running sandpile regardless of driving rate.  Changing
the rate at which sand is dropped only changes the timescales over
which the correlations are seen. This is evident in the spectra of
Fig. \ref{fig:merged_all_stretched}, where the prominent hump, whose
right side contains the $f^{-\bb}$ ($0<\beta<1$) correlated region,
moves from low frequency at low drive to high frequency at high drive.
The fact that this hump exists at all is proof of correlations in the
system, since a random superposition of pulses has a flat \fo\
spectrum at low frequencies \cite{jensen89a,kertesz90a}. This is
further confirmed by taking a activity time series from a low drive
case, where events are distinct and separated, and randomly shuffling
the events to make a new time series.  The spectrum and \rs\ of the
new time series show a \fo\ and $\hh = 0.5$ region, respectively,
indicating no correlations, whereas before both measures showed
correlated dynamics (Fig. \ref{fig:powerrsorigshufnq}). Note that this
simple experiment also tells us that it is the avalanche order in
which correlations are present, since the order of the quiet times
among them merely reflects the random character of the
drive~\cite{sanchez02b}. Therefore, the previously mentioned notions
that ``strong correlations between successive [events is] at variance
with the SOC model''~\cite{boffetta99a} and that an event ``can occur
randomly anywhere at any time and\ldots cannot `know' how large it
will become''~\cite{naturedebate,yang04a} are misleading.

Before separately discussing the physical meaning of the different
regions, it is interesting to ask ourselves why this hump and its
associated correlated region has not been discussed in the past,
giving rise to the aforementioned confusion. The only explanation we
can offer is that the time series examined were simply too short, thus
missing the timescales where correlations exist. It is instructive to
notice that in Fig. 4 of Ref.~\cite{hwakardar92a}, the power spectrum
of the low drive case exhibits only regions A, B and C, while the
spectrum for the high drive case barely extends to the beginning of
region E.  With that limited amount of dynamical information it is
reasonable that the existence of dynamical correlations was mistaken
for a consequence of the onset of avalanche overlapping.  Proper
credit, however, must indeed be given to the authors of
Ref.~\cite{hwakardar92a} for their correct interpretation of the
physical meaning of regions A, B and E. For that reason, we will just
briefly discuss them in what follows, instead emphasizing the
interpretation of regions C, D (the one that carries the signature of
long time correlations) and F.

\subsection{Regions A and B (and A/B):  Pulse Shapes}
\label{sec:partI_regions-b}

Regions A and B at low drive and region A at high drive are due to the
low level physics of the system that is manifested as the trapezoid
pulse shape of individual avalanches (see Fig.
\ref{fig:timeserieslowhigh}).  The values of the slopes and the
location of the breakpoints of these regions can in fact be predicted
by simply assuming a random superposition of trapezoids, in an
analogous way to that which was originally done in
Refs.~\cite{jensen89a,kertesz90a} and noted for the high drive case
in~\cite{hwakardar92a}. The reason why this approach works is that
regions A and B do not depend on the character of the system dynamics.
There are two high frequency regions (A and B) in the sandpile
spectrum as opposed to the single one derived in
Refs.~\cite{jensen89a,kertesz90a} because those studies assumed a
rectangular pulse shape. Such pulses have a single spectral region
while trapezoids have two. Simple tests show that replacing the
trapezoids in the sandpile activity time series by either rectangular
pulses or by spikes (`delta' functions) results in appropriate changes
to the exponents in regions A and B but in no change to regions C, D
and E. The $R/S$ analysis exhibits only one region (A/B) that ends at
a breakpoint that coincides with $T_{\mathrm{B}}^{-1}$. Therefore,
regions A and B of the power spectrum and region A/B of the \rsa\
correspond to the same low level physics of the system. In fact, note
that $H\simeq 1$ in region A/B, which implies that the value of the
breakpoint $T_{\mathrm{B}}$ is related to the time series
autocorrelation time and thus to the average avalanche duration.

Therefore, it is clear that the value of the exponents found in
regions A and B does not have a dynamical origin as has been pointed
out previously~\cite{jensen89a,kertesz90a,hwakardar92a}.  This can be
easily proved by shuffling the time series: keeping the trapezoids
intact but placing them randomly in time. This reordering necessarily
makes all dynamical correlations vanish.  It is reflected in the
spectra by the disappearance of regions D and E and the lack of change
of regions A and B (see Fig.~\ref{fig:powerrsorigshufnq}). This is an
important statement of the lowest level physics of the system. It
tells us that the high-frequency part of the spectra of systems with
different low level physics and different fundamental pulse shapes may
scale differently even if they exhibit the same low-frequency part of
the spectrum because they share the same kind of dynamics.

\subsection{Region C:  Quiet Times}
\label{sec:partI_region-c}

This is a region not previously identified in the running sandpile. In
contrast to the vanishing drive limit, quiet times (periods with no
activity) between avalanches appear in the running sandpile
\cite{sanchez02a,sanchez02b}.  In essence, quiet times are a measure
of the time that the system spends in an absorbing state.  They are
the source of the appearance of region C, which scales as $f^{\,0}$ in
the power spectrum and as $H \simeq 0.5$ in the $R/S$ analysis, both
being signatures of an uncorrelated Poisson process. These exponents
reflect the random character of the external drive that causes
avalanches to be randomly triggered.  Besides SOC models, quiet times
have been studied in other physical
systems~\cite{sanchez03b,peters02a}.

Beyond the timescale of the largest single avalanche, $T_{\mathrm{B}}
\sim \LL$, there can be a period where the correlations are dominated
by the random triggering.  The width of region C is inversely related
to the driving rate for a fixed system size. In fact, it disappears in
the higher drive cases, after the average quiet time becomes on the
order of the characteristic avalanche duration, $T_{\mathrm{B}}$.
Also, the lower the driving rate \mypol, the fewer avalanches that
occur in a given time period and the longer the system must `wait' for
enough avalanches to occur to correlate with each other.  The cutoff,
$T_{\mathrm{C}}$, is a measure of this approximate minimum time and it
scales inversely with the system time scale, $\sim \po^{-1}$.

Note that quiet times do not have a physical meaning in the zero drive
limit of any sandpile, when the addition of sand is suspended during
an active avalanche and is reinitiated immediately after it is
completed. Their presence in the running sandpile does not change the
dynamics, but allows the external drive to set a physical measure of
time, $\po^{-1}$, and to estimate the average quiet
time~\cite{sanchez02b}. The test shown in
Fig.~\ref{fig:powerrsorigshufnq} justifies this statement: the
spectrum and $R/S$ analysis of the zero drive case are identical with
those of the low drive case with quiet times {\em removed}. Note that
a low drive case must be used for this exercise, so that there are few
or no overlapping avalanches. We emphasize this to introduce the next
section, which shows that overlapping of events is not necessary to
produce correlated dynamics.

\subsection{Region D:  Dynamical Correlations}
\label{sec:partI_region-d:-soc}

\begin{figure}
  \centering
  \includegraphics[height=\onepicht]{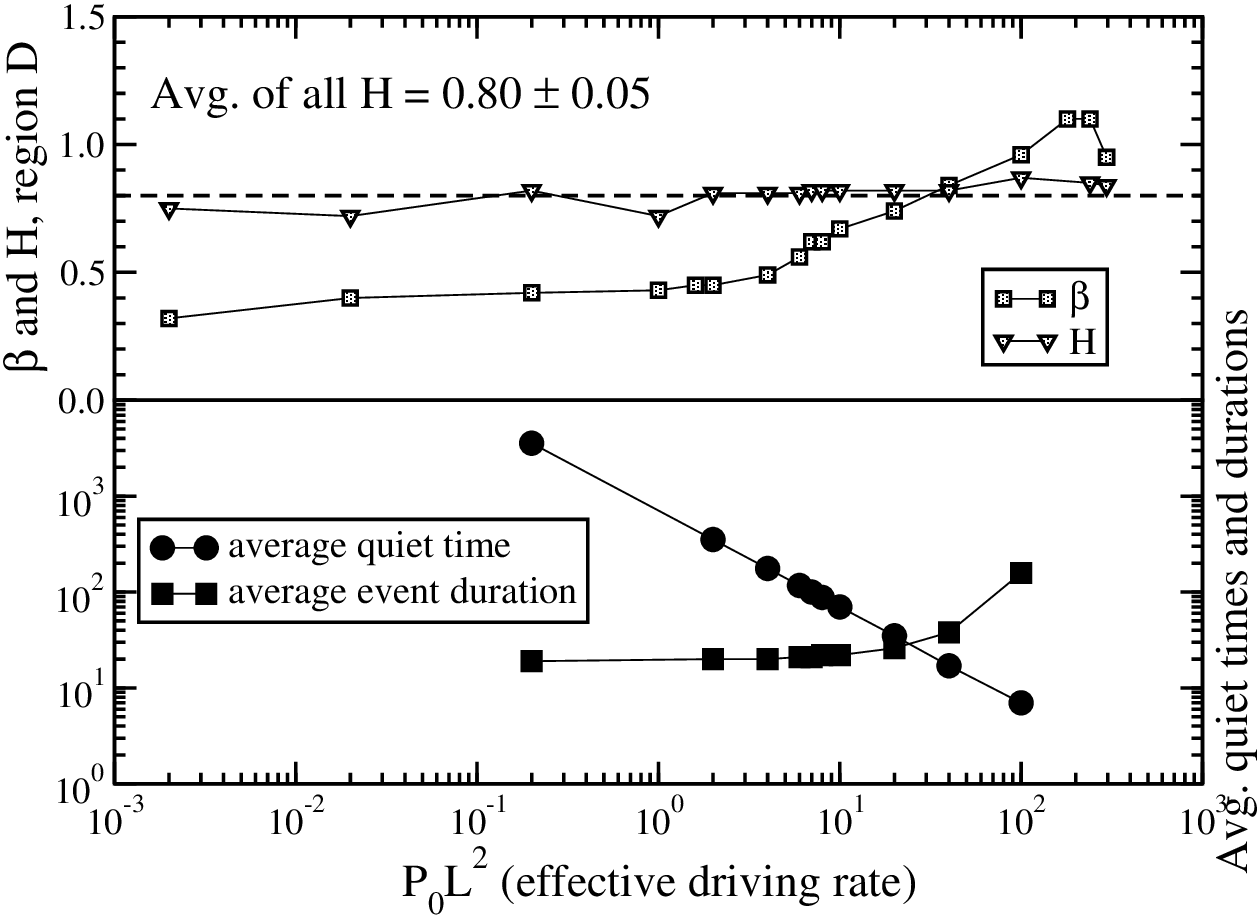}
  \caption[\hh, \bb, average quiet time and average duration versus
  five orders of magnitude of effective driving rate.]{\hh, \bb,
    average quiet time and average duration versus five orders of
    magnitude of effective driving rate.  \hh\ stays constant with
    changing driving rate while \bb\ increases as driving rate
    increases, saturating at around $\bb=1$.}
  \label{fig:constant_H_change_beta}
\end{figure}

Proving the existence of and understanding the physics of region D is
the main point of this paper. For all cases in the low drive regime,
regardless of system size, we find $H\simeq 0.8$ in region D,
indicating long time correlations. For the same timescale range, an
$f^{-\beta}$ region with $0<\beta<1$ is found in the power spectra,
again a signature of correlations. While \hh\ remains constant,
though, \bb\ changes with driving rate (see Fig.
\ref{fig:constant_H_change_beta}): it approaches $1$ at the highest
drives and decreases as drive strength is reduced.

It is easy to show that the correlations must arise from interactions
among distinct events and not from overlapping of events. The first
clue is that this region exists on timescales far greater than the
maximum duration of an individual avalanche ($T_{\mathrm{B}}$).  Since
the activity time series consists of a succession of avalanches separated
by quiet times and there is no (or very little) overlapping in the low
drive regime, the correlations must be due to the specific order of
events, chosen by the system dynamics in spite of the randomness of
the drive.  This is confirmed by the random shuffling of the activity
series mentioned when discussing regions A and B (see Fig.
\ref{fig:powerrsorigshufnq}). Regions C, D, E and F collapse into a
single uncorrelated region for all time scales beyond
$T_{\mathrm{B}}$, where the shuffled series has a $f^{\,0}$ spectrum
down to the lowest frequencies and a Hurst exponent $H \simeq 0.5$ to
the longest time scales.

A second confirmation that the correlations are established among the
separate events is that quiet times have no role in the correlations.
This is confirmed by the test already mentioned when discussing region
C: removing all quiet times from the activity data while keeping the
event order unchanged.  As seen in Fig.  \ref{fig:powerrsorigshufnq},
region C disappears: the $H \simeq 0.5$ region is replaced by a $H
\simeq 0.8$ region, which means that the $H \simeq 0.8$ region D of
the original data has moved to shorter timescales due to the removal
of quiet times and the concomitant shortening of the series. But the
correlations among events remain the same since neither their order
nor their sizes have been changed.

\begin{figure*}
  \centering
  \begin{tabular}{cc}
    \includegraphics[width=\twopicwt]{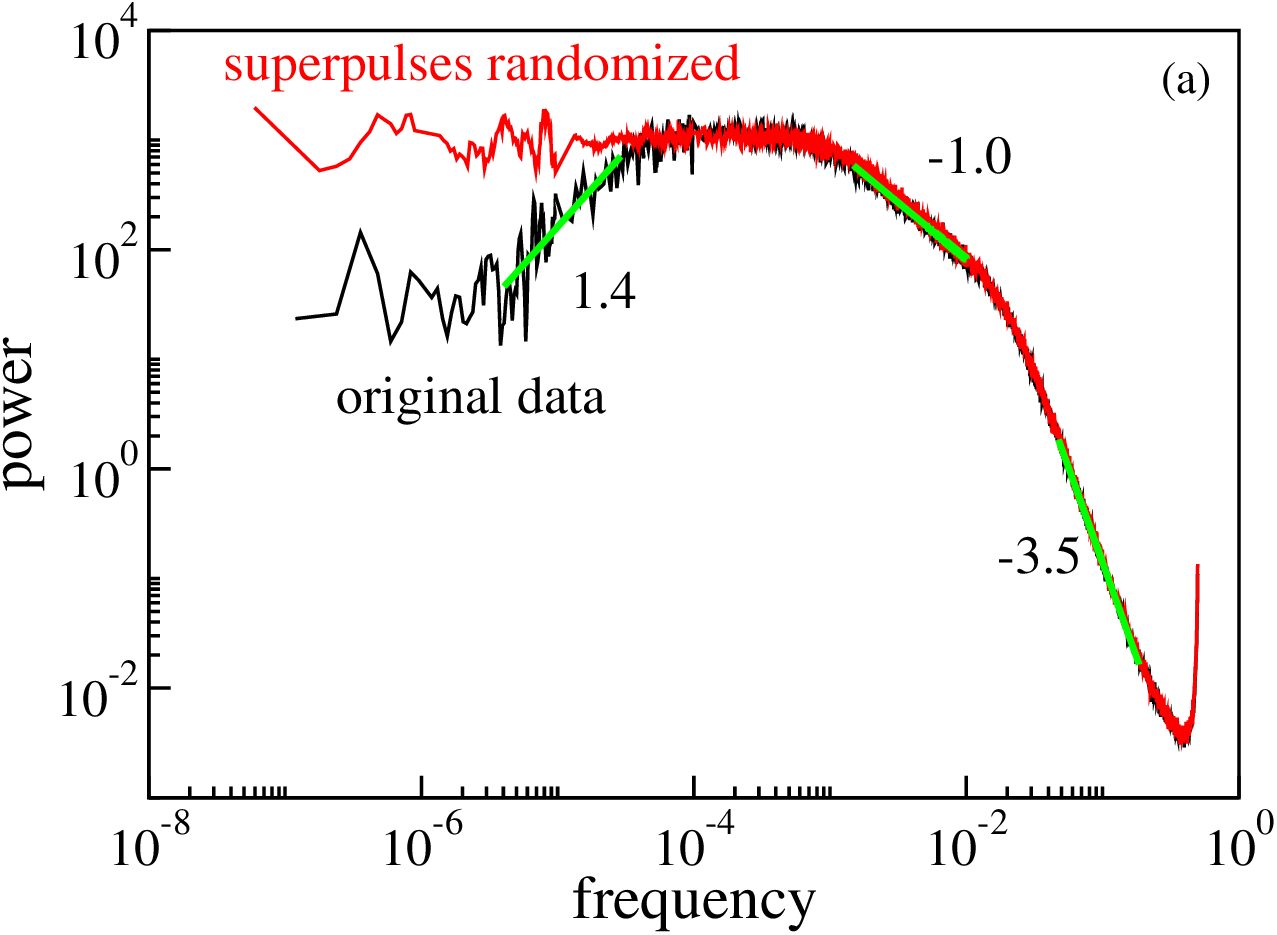} &
    \includegraphics[width=\twopicwt]{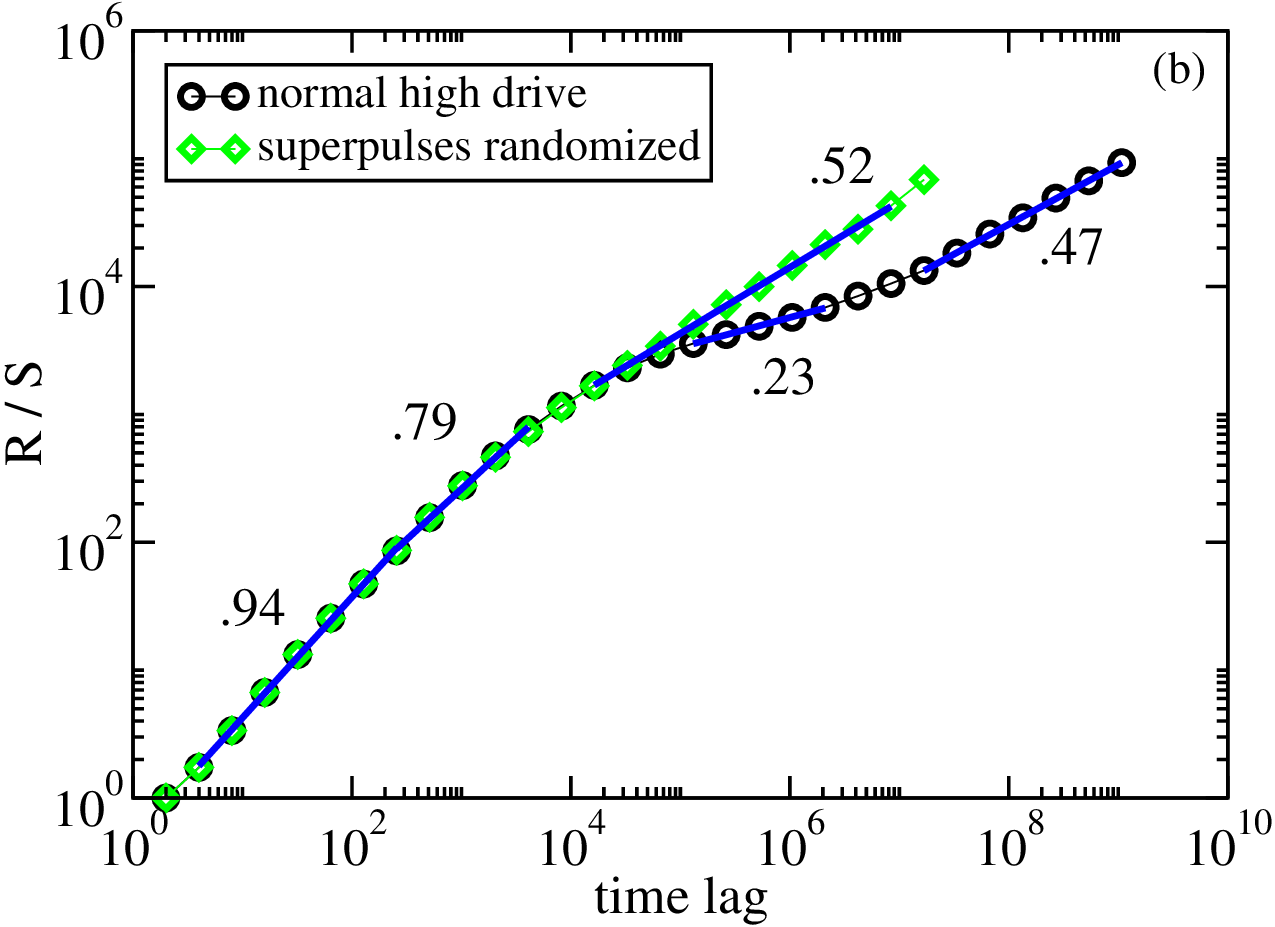} \\
  \end{tabular}
  \caption{(Color online.) Power spectra and \rs\ for high drive with
    superpulses changed.  Numbers shown are \bb\ for spectra and \hh\
    for \rs.}
  \label{fig:spectrarshighsuperpulse}
\end{figure*}

The same value of $H \simeq 0.8$ is found for this region regardless
of system size or driving rate, even in the vanishing drive limit in
which SOC is strictly defined~\cite{dickman98a}. This simple fact
suggests that the correlations present at the SOC critical point do
persist almost without distortion at finite drives. In fact, this is
still the case even at the highest drives, when avalanches do overlap
and form superpulses. A superpulse is defined as the structure between
successive quiet times in a high drive activity time series (Fig.
\ref{fig:timeserieslowhigh}(b)). Note that quiet times in the high
drive cases are very small, on average. By randomly shuffling the
superpulses of a high drive time series, we see that the correlations
that lead to $\hh \simeq 0.8$ are on time scales shorter than the
average superpulse.  Beyond these time scales, $\bb \simeq 0$ and $\hh
\simeq 0.5$, signatures of uncorrelated data (Fig.
\ref{fig:spectrarshighsuperpulse}).

Understanding the power spectrum of region D is not as simple as for
the $R/S$ analysis.  The slope of this region, $\beta$, changes with
driving rate and with system size. $1/f$ regions are only obtained for
high drives (or by removing the quiet times of low drive cases, which
makes the system approach the zero drive limit as previously
discussed). This implies that a $1/f$ region is not a necessary
signature of SOC dynamics at finite drives, since correlations are
still present, as indicated by the Hurst exponent remaining constant
at $H \simeq 0.8$. In this sense, $R/S$ analysis is a more robust
measure of the degree of correlations among events than is the power
spectrum.

\begin{figure}
  \centering
  \includegraphics[height=\onepicht]{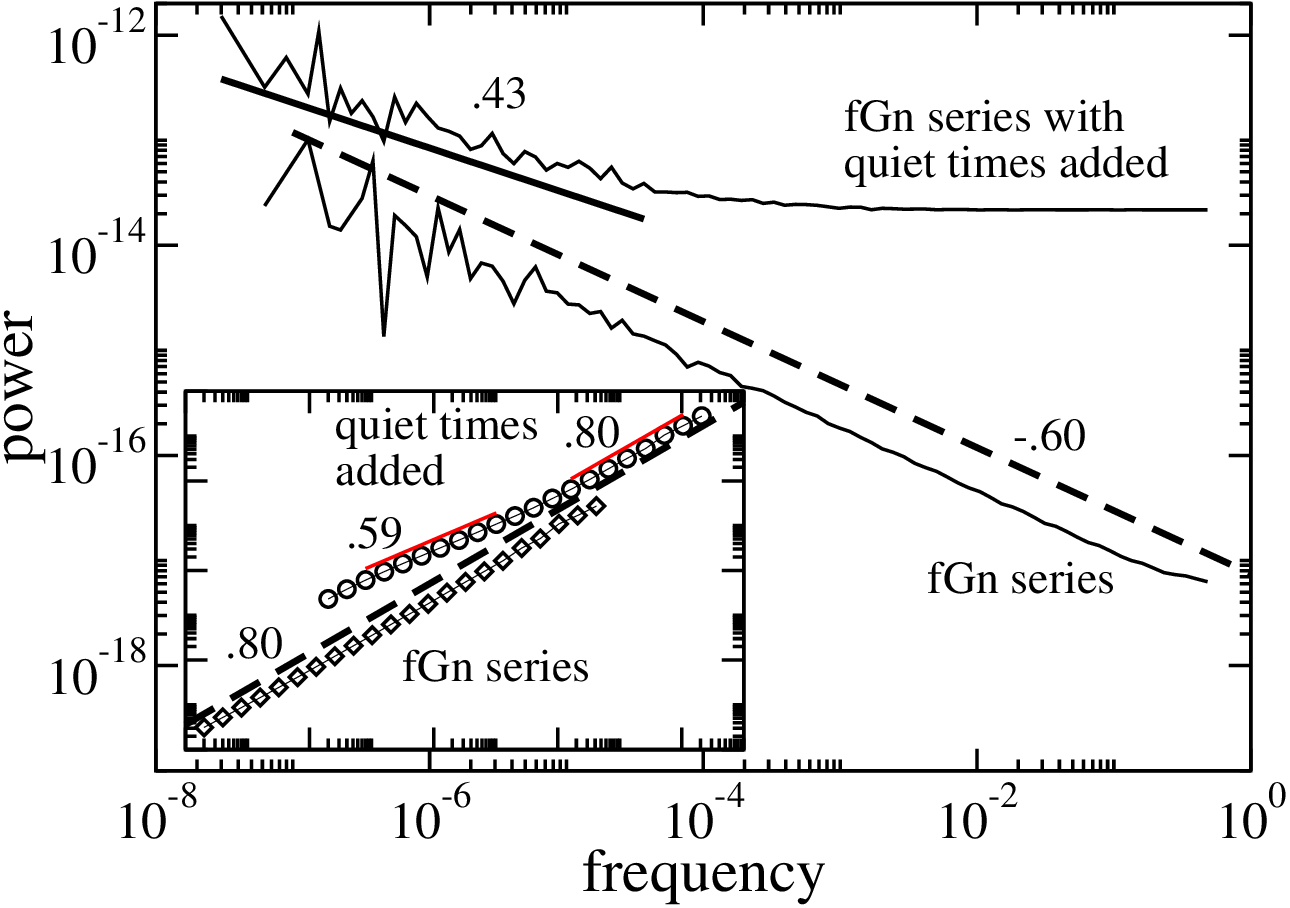}
  \caption[Power spectra and (inset) \rsa\ for fractional Gaussian
  noise of $\hh = 0.8$ with and without quiet times added.]{Power
    spectra and (inset) \rsa\ for fractional Gaussian noise of $\hh =
    0.8$ with and without quiet times added.  Numbers shown are \bb\
    for spectra and \hh\ for \rs.}
  \label{fig:fGn__8_power_rs_PII}
\end{figure}

The question, however, remains open about why \bb\ changes as region D
is pushed towards different timescales, and why \bb\ saturates at $1$
and not another value. We do not know the answer, but we can say a
couple of things.  First, it appears that \ff\ is due more to a lack
of quiet times and distinct pulse shapes than to overlapping, at least
in the sandpile model.  We showed in Fig.  \ref{fig:powerrsorigshufnq}
that removing all quiet times from between separate events of a low
drive activity time series produces a spectrum with a region D where
$\bb = 1$. In that case, there is no overlapping of events by design.
We can also discard the need for hidden additional correlations being
established dynamically, since the same change in \bb\ is observed
when inserting randomly-distributed quiet times into an artificial
time series of fractional Gaussian noise (fGn) \cite{mandel1999a} with
an arbitrary Hurst exponent $\hh = 0.8$ (see
Fig.~\ref{fig:fGn__8_power_rs_PII}). We inserted quiet times with an
average approximately the same as the average of the original fGn
series ($\simeq 20$). The spectrum and \rsa\ are shown for the
original and modified series. The original fGn series follows \bbh,
with $\hh \simeq 0.8$ and $\bb \simeq 0.6$. When quiet times are added
between each point of the original series a region of $\hh \simeq 0.5$
appears up to a certain time lag. Beyond this time lag, $\hh \simeq
0.8$ as expected since the correlations among the data have not
changed. In the spectrum, $\bb \simeq 0$ down to the inverse of the
same time lag and then $\bb \simeq 0.43$.  This constant \hh\ and
changing \bb\ is the same effect seen in the sandpile data.

\subsection{Region E: Global Discharge Region}
\label{sec:partI_region-e-low}

Region E is the discharge event region, already well-studied by
Ref.~\cite{hwakardar92a} for high drive, and a consequence of the
finite system size. The finite sandpile is always being driven towards
a globally critical slope. Once all or almost all cells have slope
$\geq Z_{\mathrm{crit}} - \nf$, a keystone toppling, usually near the
bottom, will produce a system-size avalanche or a rapid succession of
smaller avalanches that removes enough sand so that the slopes at all
sites are reduced to much less than critical ($< Z_{\mathrm{crit}} -
N_{\mathrm{f}}$).  Once such a large event occurs, it is unlikely that
one will happen again for a long time. Hence, large events are
anticorrelated and have the signatures of anticorrelated dynamics,
$\beta < 0$ in the spectrum and $H < 0.5$ in $R/S$ analysis. The same
process drives both the low and high drive cases.

\subsection{Region F: Beyond Dynamics}

Beyond breakpoint $T_{\mathrm{F}}^{-1}$, at the lowest frequencies,
the spectrum is flat, which we conjecture to be a reflection of the
random drive of the system on the longest time scales. Note that all
dynamics is on time scales much smaller than those in this region.
Correlations are eventually erased by the system-wide events and
shuffled by the random drive at time scales longer than
$T_{\mathrm{F}}$.  Region F extends to infinitely low frequencies;
there are no further dynamical regions below it.

\section{Relation between system size and driving rate}
\label{sec:size-drive}

We return to Figure~\ref{fig:flipsdiffLpowerrs} to discuss the effect
of changing system size.  It is reasonable to increase the level of
forcing---practically, the amount of avalanche overlap---in the system
by increasing the driving rate \po.  This means that more grains per
time step will fall on the sandpile and triggering time will be
reduced.  It turns out that keeping \po\ fixed while increasing system
size \LL\ has the same effect because avalanches last longer in the
larger system.  In both of these cases, \mypol\ has increased, causing
trigger time to decrease.  But since system size \LL\ has remained
constant, the ratio of average avalanche duration to trigger time has
increased further, producing more overlap and stronger effective
forcing.

But the combination \mypol\ does not by itself give enough information
about the character of the drive as represented by the amount of
avalanche overlap.  Imagine increasing \LL\ while decreasing \po\ so
that \mypol\ and, therefore, average trigger time remain constant.
Though trigger time remains constant, the ratio of average avalanche
duration to trigger time has again increased because of the increase
in system size.  This process, then, also effectively increases the
driving rate in the larger system by producing more overlap among the
individual avalanches that now last longer.  This is true even though
the original and the larger system both have the same \mypol\ and,
therefore, the same flux into the whole system and out of the bottom
of the sandpile.

In terms of the effective driving rate discussed in Section
\ref{sec:model}, \je\ is larger for the larger system when \mypol\ is
the same in both.  This feature can be exploited to produce a plot
qualitatively similar to Fig.  \ref{fig:merged_all_stretched}.  In
that figure, the driving rate \mypol\ of the sandpile increases from
top to bottom for the spectra shown while system size \LL\ remains
constant.  If instead, one started with the spectrum of a small system
at the top of the figure and increased system size while keeping
\mypol\ constant, then the \emph{same figure would be obtained}.  In
both cases, \mypoll\ increases from top to bottom.  An application of
this is that perhaps one wants to study a physical system that is
suspected of being described by SOC and/or the running sandpile.  If
experimenters cannot change the system size (like, for instance, that
of the magnetosphere) then perhaps periods of different observed
driving rates could be used as a proxy for different system sizes.

\begin{figure}
  \centering
  \includegraphics[height=\onepicht]{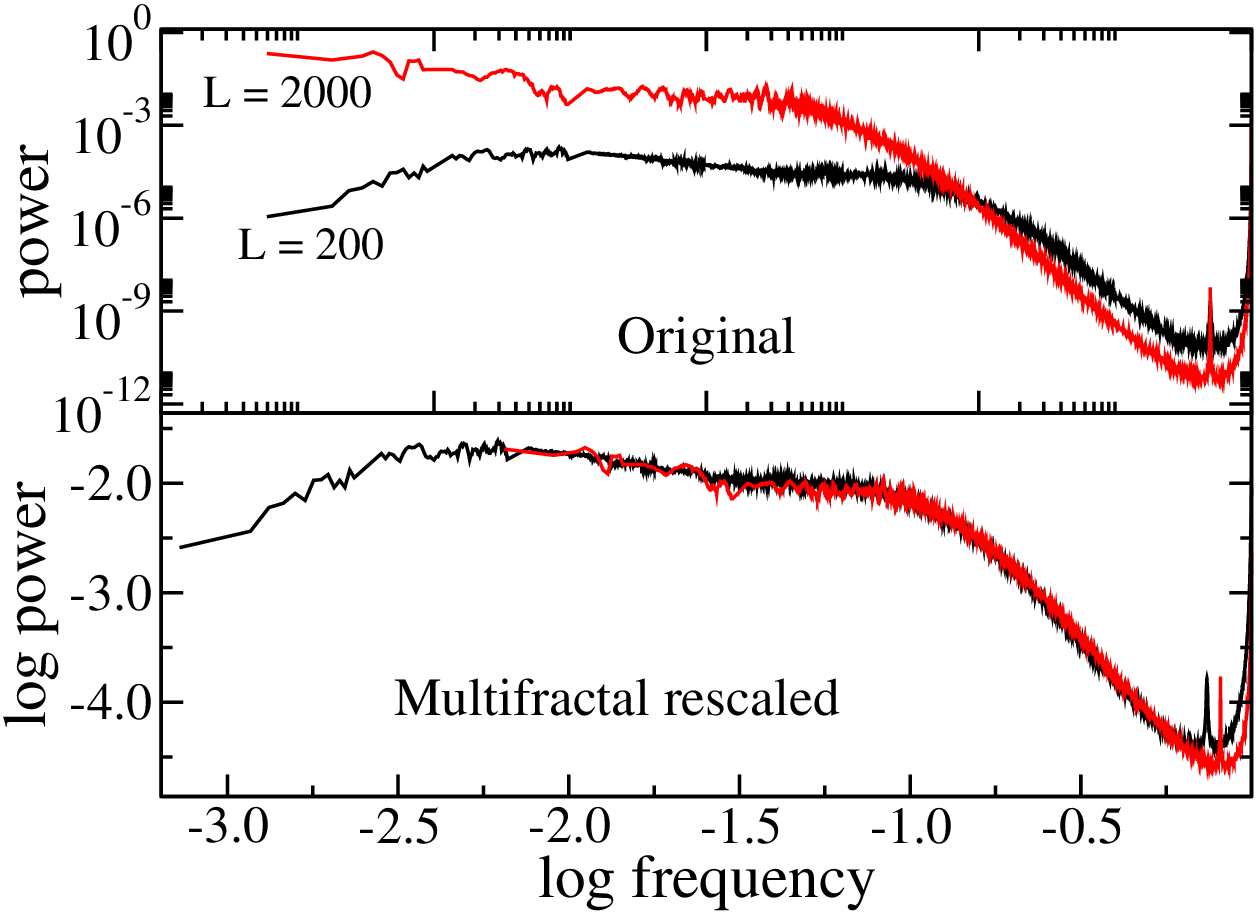}
  \caption{(Color online.) Rescaling of power spectra of two different
    systems with $\mypoll = 2.0$, one with $\LL = 200$ and one with
    $\LL = 2000$, using the rescaling law $y_0 \log_{10} S (f) = g
    \left(x_0 log_{10} (f / x_1) \right)$, where $S(f)$ is the power
    spectrum and $g(f)$ is a scaling function. Different sets of
    parameters $[x_0, x_1, y_0]$ are needed for each spectrum so this
    is not the same as, for instance, the multifractal rescaling of
    avalanche size PDFs of \cite{kadanoff89a}, where the same value of
    each parameter is used for different system sizes.}
  \label{fig:pollrescale}
\end{figure}

Since $\je$ is the correct quantity to use to compare systems in the
same drive regime but where the individual parameters are different,
this has implications, for example, in investigations of finite size
and/or multifractal scaling.  As an illustration, Fig.
\ref{fig:pollrescale} shows justification for comparing systems with
the same effective driving rate, in this case with $\mypoll = 2.0$.
The spectra of systems with different driving rates \mypol\ and system
sizes \LL\ but with the same \mypoll\ can be rescaled to lie on top of
each other.  The spectra of systems with the same driving rate \mypol\
but with different system size \LL\ cannot be rescaled to lie on top
of each other.

\renewcommand{\cap}{(Color online.) Rescaling of PDF of avalanche
  durations for two system sizes, $\LL = 200$ (solid line) and $\LL =
  2000$ (dashed line), and a shared effective driving rate $\mypoll =
  2.0$.  The PDFs have been rescaled by $\LL^{\alpha}$ where $\alpha =
  0.34$.}
\begin{figure}
  \centering
  \includegraphics[height=\onepicht]{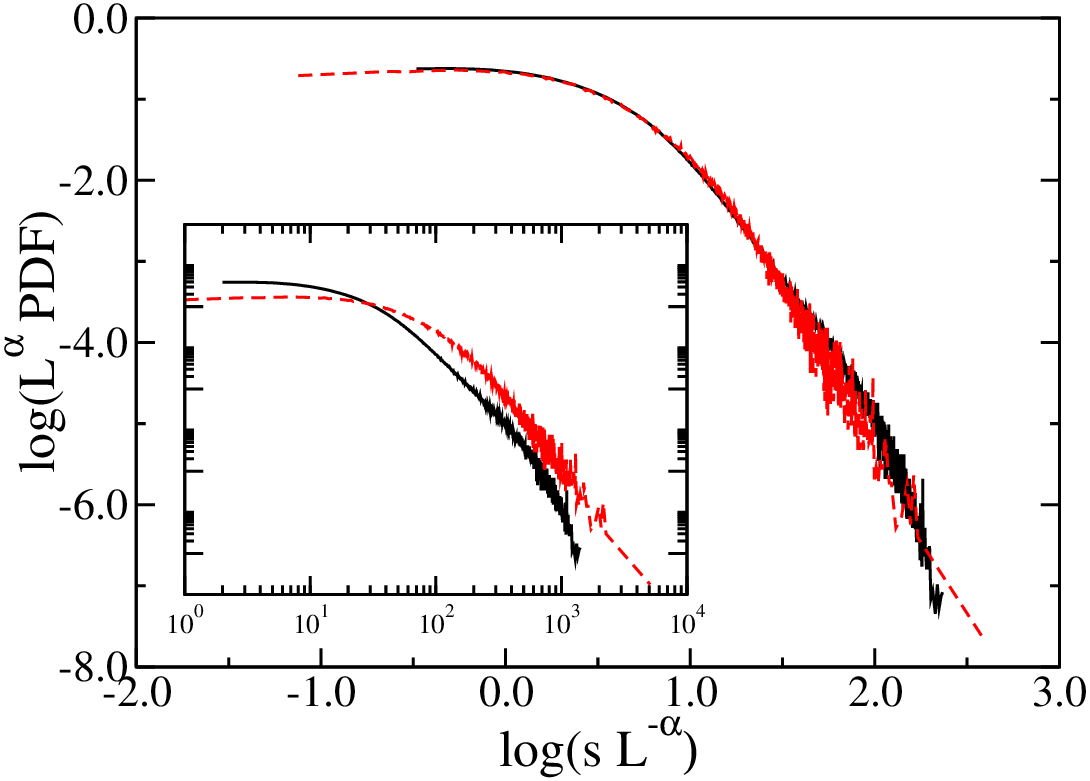}
  \caption{\cap}
  \label{fig:pdfrescale}
\end{figure}

Finally, we show in Fig.~\ref{fig:pdfrescale} the PDFs of avalanche
durations from two different system sizes that have been rescaled by a
factor of $\LL^{-\alpha}$ with $\alpha = 0.34$.  The rescaling has
been done for a medium driving rate of $\mypoll = 2.0$ so that there
is some avalanche overlapping.  This, coupled with the rescaling of
the spectra, shows that the running sandpile exhibits critical
dynamics similar to that seen in the sandpile with vanishing drive
\cite{kadanoff89a}.

\section{Conclusions}
\label{sec:conclusions}

We have analyzed the one dimensional directed running sandpile model
for five orders of magnitude of driving rate and for almost three
orders of magnitude of system size and have shown that the same
dynamical correlations that are present in the SOC zero-drive limit
persist (and produce nontrivial signatures in the power spectra andWe
show through rescaling of power spectra and avalanche size PDFs that
the running sandpile exhibits SOC in the same way that the vanishing
drive sandpile does.  $R/S$ analysis of activity time series) at
finite drives.  Our results warn that a time series from {\em any}
system, for which SOC may provide a plausible description of the
dynamics, may be too short to see these correlation signatures and
could thus be mistaken for a simple random time series.  We show
through rescaling of power spectra and avalanche size PDFs that the
running sandpile exhibits SOC in the same way that the vanishing drive
sandpile does.W

The reason why these correlations persist away from the SOC critical
point at zero drive can be understood when discussing the physical
mechanism through which the sandpile establishes dynamical
correlations. It is well known that memory in the sandpile is retained
in the heights (and, therefore, the local gradients) of each cell.
This is, for a physical system, the instantaneous system state. When
an avalanche ends, the cells at the endpoints of its active zone are
closer to critical and are more likely to be the initiation point of
another avalanche.  This is the basis for memory in the system and
explains why the drive strength does not affect dynamical
correlations, except for pushing them to longer or shorter time
scales: {\em regardless of how long the intervening quiet time is
  between two events, they are correlated in the same way.} In
reference to the introduction, this is how an event can `know' how
large it will become---it can grow only as large as the size of the
local neighborhood of marginal cells. Also, this is why an event of
any size {\em cannot} simply occur at any time or place in the system.
Only the time of the triggering of an event is random due to the
random external forcing.  Before that trigger, though, the size and
location of the event is predetermined by the state of the local
neighborhood, like an earthquake fault ready to slip.  Therefore, the
same memory mechanism that is the basis for SOC at zero drive is still
present in the gradients regardless of how slowly or quickly sand is
added, as long as the system is not overdriven.

As we have shown, region D is the only region that reflects the
underlying zero-drive SOC dynamics at finite drive: {\em on the time
  scales in this region, the spectral and \rs\ signatures reflect only
  long time correlations and nothing about pulse shape, quiet times,
  random superpositions, overlapping of pulses or system size.} In
fact, in the thermodynamic limit in which the system size becomes
increasingly large, we must simultaneously allow $P_0\rightarrow 0$ to
keep $\je = P_0L^2$ constant, which pushes the system to the SOC
critical point~\cite{dickman98a,dickman00a,vespignani00a}. In that
case, region D would extend to lower and lower frequencies.

Finally, another interesting result derived from this work is that the
relation \bbh\, derived in Ref.~\cite{mandel2002a} for fGn series and
often accepted as the correct relationship between \bb\ and \hh\
regardless of the system, does not hold in general.  While this fact
is known from general theory~\cite{mandel82a,mandel2002a}, the
sandpile results provide a concrete example where the relation does
not hold.  This implies that the random drive \po\ of the system for
which \bbh\ \emph{does hold} is filtered by the sandpile so that \bb\
and \hh\ no longer have such a simple relation.  Note that the
changing \bb\ and constant \hh\ of the correlated region D imply that
the mechanism that produces \ff\ at the high drive produces non-\ff\
at lower drive, so that `\ff\ is not always \ff' and smaller values of
\bb\ are not necessarily any less `special' than $\bb = 1$.

\ack Many thanks to H.~J.~Fletcher, N.~W.~Watkins and B.~T.~Werner for
support and suggestions.  Support from DOE under grants
DE-FG03-99ER54551 and DE-FG03-00ER54599 (a young investigator award)
and NSF under grant ECS-0085647 are gratefully acknowledged.


\end{document}